\documentclass[a4paper,12pt]{article}
\usepackage{amsmath,amssymb,amsthm,mathrsfs,graphicx,float,colordvi,url,wrapfig,color,ulem}
\usepackage[all]{xy} 
\usepackage{tikz-cd} 

\def\green{\textcolor{black}} 

\def\Det{{\rm Det}} 

\begin{document}

\begin{titlepage}

\vspace*{5mm}

\begin{center}
{\bf \large
Josephson junction formed in the wormhole space time
}
\\
\vspace*{2.5mm} 
{\bf  \large
from the analysis for the critical temperature of BEC
}

\vspace*{8mm}

\normalsize
{\large Shingo Takeuchi}

\vspace*{6mm} 

\textit{{}$^\dagger$Phenikaa Institute for Advanced Study and Faculty of Basic Science,}\\
\vspace*{0.10 cm}
\textit{Phenikaa University, Hanoi 100000, Vietnam}\\ 

\vspace*{2.0 mm}

\end{center}

\vspace*{5mm}
\begin{abstract}
In this study, considering some gas in the Morris-Thorne traversable wormhole space time, 
we analyze the critical temperature of the Bose-Einstein condensate in the vicinity of 
its throat. As a result, we obtain the result that it is zero. Then, from this result, we point 
out that an analogous state to the Josephson junction is always formed at any temperatures 
in the vicinity of its throat. This would be interesting as a gravitational phenomenology. 
\end{abstract}
\end{titlepage} 

\allowdisplaybreaks 
\section{Introduction} 
\label{jeloiw} 

The issues concerning the wormhole space times as the solution and the phenomena of these 
are one of the problems that have been investigated so much until now. Since this paper will 
treat a phenomenological issue in the wormhole space times, we refer some phenomenological 
works on the wormhole space times; 
gravitational lensing \cite{Abe:2010ap,Nakajima:2012pu,Yoo:2013cia,Tsukamoto:2016zdu,Tsukamoto:2016qro,Tsukamoto:2017hva,Nandi:2018mzm,Ono:2018ybw}, 
shadows \cite{Nedkova:2013msa,Ohgami:2015nra,Ohgami:2016iqm}, 
observation \cite{Dai:2019mse}, 
Casimir effect \cite{Khabibullin:2005ad}, 
teleportation \cite{Susskind:2017nto}, 
collision of two particles \cite{Tsukamoto:2014swa}, 
and creation of a traversable wormhole \cite{Horowitz:2019hgb}. 

However, in these years, studies regarding wormhole space times are performed so energetically. 
Its reason would be that recently it has turned out that 
the Einstein-Rosen bridge (ER bridge) \cite{Einstein:1935tc} gives 
some kinds of the EPR pair \cite{Maldacena:2013xja,Jensen:2013ora,Jensen:2014bpa,Bao:2015nca}, 
which plays very important role in the literature of the information paradox \cite{Susskind:2013lpa}. 
It would be also its reason that  there are common features and behaviors between AdS$_2$ wormholes 
and the SYK models, from which currently we can perform various interesting studies 
\cite{Maldacena:2018lmt,Garcia-Garcia:2019poj,Chen:2019qqe,Maldacena:2019ufo}. 

Also, there are studies to construct the graphene wormhole in the material physics 
from some brane configurations in the superstring theory \cite{Sepehri:2017dky,Capozziello:2018mqy}. 
From these studies, Chern-Simon current in the graphene wormhole is studied in \cite{Capozziello:2020ncr}. 
\newline

In this study, considering the Morris-Thorne traversable wormhole (traversable wormhole) \cite{Morris:1988cz}, 
we consider some situation that some gas fills the whole space time of that, where it is assumed 
Bose-Einstein condensate (BEC) can be formed at some temperature in this gas. Then, we point 
out that an analogous state to the Josephson junction is always formed at any temperatures except 
for zero in the vicinity of its throat. 

For this, we first analyze the critical Unruh temperature of the gas regarding BEC in the Rindler, 
then check that it can agree to the critical temperature obtained in the flat Euclid space.

The motivation of this is to check the rightness of our analytical method to obtain the critical temperature of BEC 
in curved space times. Actually, our analytical method in the curved space time is different from the usual one 
in the flat Euclid space \cite{Kapusta:2006pm} to treat curved space times. However, since the space time 
for the accelerated system can be regarded as the flat space with the temperature same with the Unruh temperature 
if Euclideanized (see the text given under (\ref{lqvue})), 
the critical Unruh temperature obtained in the Rindler space should agree to that in the flat Euclid space. 

Then, considering the same gas in the traversable wormhole, we analyze its critical temperature concerning BEC 
in the vicinity of its throat. 

Here, it is known that if the number of its spatial dimensions is 2 or less in the flat Euclid space, 
the effective potential always gets diverged for the contribution of the 
infrared region and BEC is not formed\footnote{
In the $1+d$ dimensional flat Euclid space, 
the particle density at the critical point of BEC is given as 
$\displaystyle 
d=\int \frac{d^dk}{(2\pi)^d}
(\frac{1}{e^{\beta(\sqrt{k^2-m^2}-\mu)}-1}-\frac{1}{e^{\beta(\sqrt{k^2-m^2}+\mu)}-1})$ in \cite{Kapusta:2006pm},
where $\mu=m$. Then we can see this is diverged for the contribution of small $k$, if $d=1$ or $2$.
At this time, the critical temperature becomes $0$ \cite{becd12}, which means the state is normal at any temperature.
}. Then, since the spatial part of the vicinity of the throat 
in the traversable wormhole space time becomes effectively 1 dimensional 
(as for this point, see (\ref{qdjes}), then just take the limit: $r \to v$). 
we can expect that the critical temperature of BEC in the vicinity of the throat is always zero. 
(Of course, what the dimension effectively becomes 1 depends on the coordinate system we take. 
Therefore, surely we could expect the zero critical temperature in the vicinity of the throat if the 
dimension becomes 1, however it would be considered it is not its essential reason. 
We will give comment on our result in the end of Sec.\ref{desiuw} from this point of view.) 

Then, 
\begin{itemize}
\item
If the critical temperature is always zero in the vicinity of the throat, 
since the throat exists in the form to separate the wormhole space time 
from the another side of the wormhole space time, the normal state also 
appears in the form to separate the wormhole space time 
from the another side of the wormhole space time like Fig.1. 
\vspace{-10mm}
\begin{figure}[h!!!!!] 
\label{dsqaui} 
\vspace{0mm} 
\hspace{0mm}
\begin{center}
\includegraphics[clip,width=4.0cm,angle=-90]{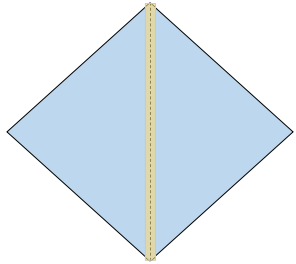} 
\end{center} 
\vspace{0mm} 
\caption{
Penrose diagram for the ($t$, $r$) part of the traversable wormhole space time. 
The dotted line represents the throat of the wormhole. 
The yellow and blue parts represent the normal and superconductor states we expect to appear, respectively. 
As the normal state is appearing in the form to separate the space time, 
we can expect that an analogous state to the Josephson junction is formed in the vicinity of the throat.}
\end{figure} 
\item
The far region of the wormhole space time is asymptotically flat space time. 
Therefore, the critical temperature of BEC in the far region is given by the 
one calculated in the flat space time.
\item
Then, considering by extrapolating between the results in the throat region and the far region, 
we can obtain an expectation that an analogous state to the Josephson junction is formed 
at any temperatures in the vicinity of the throat. 
\end{itemize}

Since this could be considered as a gravitational phenomenology, this would be interesting.
This would be also interesting because we can consider a possibility that the Josephson current flows 
from the one side of the wormhole to another side. Of course, the wave function of the current may be 
damped when it tunnels, and the Josephson current does not exist in practice. 
Currently we could not say any explicit things on this point from the analysis in this study, which is a future work. 
We discuss this in Sec.\ref{vewrc}.
\newline

The wormholes we can consider would be the ER bridge and the traversable wormhole space time. 
Then, since the ER bridge is, briefly saying, given by patching two Schwarzschild black holes together, 
the Josephson current will flow out from the event horizon. 
Hence,  it would be lightlike, and could not be observed at any points in the outside of the event horizon 
(in the regions I and III in the Penrose diagram) except for the future timelike infinity. 
Therefore, it might not be suitable to consider the ER bridge if we consider about the Josephson current.  


The traversable wormhole is free from this problem as we can depict some timelike flow of the current 
going across the throat part in Fig.\ref{dsqaui}. 
However it needs the exotic matter (some matter to violate the energy condition) \cite{Visser:1995cc} 
such that the traversable wormhole space time can satisfy the Einstein equation, which we explicitly show in Sec.\ref{vvbep}. 
This is a critical problem as it means that the traversable wormhole space time is not realizable realistically. 
For some recent studies on the exotic matter, 
see \cite{Kuhfittig:2003pu,Kanti:2011jz,Kanti:2011yv,Moraes:2017dbs,Antoniou:2019awm,Samanta:2019swa,Jusufi:2020yus,Sharma:2021egn}.
In this study, we consider the traversable wormhole upon knowing this problem. 
If we care this problem in this study, we could mention that 
it appears for some reason and can exist for some time, 
then this study is the one for that time. 
\newline

We mention the organization of this paper. 
We analyze the critical Unruh temperature of the gas regarding BEC 
in the Rindler space (Sec.\ref{qwebpr})
in the Euclid space (Sec.\ref{qbele})
in the traversable wormhole space time (Sec.\ref{bsbrtw}). 
Then in Sec.\ref{vewrc}, we point out that an analogous state to the Josephson junction is formed in the vicinity of its throat. 
In Sec.\ref{vewrc}, we obtain the phase structure for the BEC/normal states transition in the ER bridge. 
In Appendix.\ref{dcmqt}, we note some points in the mechanism for the formation of BEC in this study.

\section{Critical temperature of BEC in accelerating system} 
\label{qwebpr} 

\subsection{Rindler coordinates and Unruh temperature} 
\label{ksdvb} 

As a theoretical model, we consider some gas that entirely fills the Minkowski spacetime and performs an uniformly accelerated motion 
with an acceleration $a$ for one direction. 
Based on (\ref{cpqaed})
\footnote{
Solving $\frac{d}{dt} ( \frac{m v}{\sqrt{1-\beta^2}})= m a$ with the condition $v=0$ at $t=0$ and the general relation between 
the Minkowski time and proper time: $\tau=\int_0^t dt \sqrt{1-\beta^2}$, the trajectory of an accelerated motion can be 
obtained as
\begin{align}\label{cpqaed}
t = \frac{c}{a}\sinh \frac{a\tau}{c}, \quad
x = \frac{c^2}{a} \,(\cosh \frac{a\tau}{c}-1),
\end{align} 
where we can check with the $t$ and $x$ above, we can obtain $ ( \frac{m v}{\sqrt{1-\beta^2}})= m a t$. 
Toward this, there are several ways for how to take the coordinate system along an accelerated 
motion \cite{wikils}.
}, 
we denote the coordinates of the gas as
\begin{align}\label{vrbwo}
t = \frac{\xi}{a}\sinh a\zeta, \quad
x = \frac{\xi}{a} \cosh a\zeta,
\end{align}
which are a kind of the Rindler coordinates, and at $\xi=1$, $\zeta$ becomes the proper time 
of the object performing the uniformly accelerated motion $a$. 
Then, the squared line element can be given as follows:
\begin{align}\label{astew}
ds^2 =& \, \xi^2 d\zeta^2 -a^{-2} d\xi^2 - dx_\perp^2, \nonumber\\
       =& \, \rho^2 d\tau^2 -d\rho^2 - dx_\perp^2,
\end{align}
where $\rho \equiv a^{-1}\xi$ and $\tau \equiv a \zeta$. 

Here, $(y,z)$ are common with those in the Minkowski coordinates and we 
use the notation $x_\perp \equiv (y,z)$ in what follows. Note that we take 
not a sphere coordinates but a plane coordinates for the $x_\perp$-direction. 

Euclideanize the coordinate system with $\tau \to i\tau$, and regard it as periodic. 
Then, from the no conical singularity condition: $\frac{\textrm{circumferential length}}{\textrm{radius}}\big|_{\rho=0}=2\pi$, 
the period in the $\zeta$-direction can be determined as $2\pi/a$.  
Therefore, the gas in the Rindler coordinate system with the proper time $\zeta$ 
can be would be at the following temperature: 
\begin{align}\label{lqvue}
T_U= a/2\pi=1/2\pi \rho,
\end{align} 
which agrees to the Unruh temperature.

Here, in (\ref{astew}), when we put $\rho$ to $1$, $d\rho^2$ vanishes. 
However, let us define $d\rho^2$ as the squared line element perpendicular to the original $\rho$-direction. 
Then, since the period of $\tau$-direction is $2\pi$, the period in terms of $\zeta$ becomes $2\pi/a$. 
Hence, when we put $\rho$ to $1$, (\ref{astew}) can be regarded as the $D=1+3$ flat Euclid space 
with the Euclid finite temperature $T_E=2\pi/a$ for the one with the proper time $\zeta$. 
Therefore, a space time for the accelerated system can be considered to be equivalent to the flat space with the temperature same 
with the Unruh temperature if Euclideanized, so that the critical Unruh temperature obtained in the Rindler space 
would be considered to agree to that in the flat Euclid space. 

We can see from (\ref{lqvue}) that 
locating at different $\rho$ means having different Unruh temperature. 
Therefore, we should be careful when we perform the $\rho$-integral. 
For this point, we will consider that the Rindler coordinates are individually 
applied for each particle comprising the gas, and all the particles will 
have the same value with regard to $\rho$. 
As for the treatment of $\rho$ in the analysis in this study, we will 
consider the effective potential at some $\rho$ as in (\ref{W_Dpm})\footnote{
In \cite{wikils}, there is some comment for another problem in order for an 
object with some volume performing an accelerating motion to keep its shape.}. 

\subsection{Hamiltonian in finite density, probability amplitude and Euclideanization}
\label{bsow} 

We start with the following Lagrangian density for the complex scalar field 
for the particles comprising the gas that we mention in Sec.\ref{ksdvb} that fills the whole space: 
\begin{align}\label{sfdkdf} 
{\cal L} = \, g^{\mu\nu} \partial_\mu \phi^* \partial_\nu \phi - m^2 \phi^* \phi, 
\end{align}
where $\phi=\frac{1}{\sqrt{2}}(\phi_1+i \phi_2)$ ($\phi_{1,2}$ are real scalar fields), 
Indices $\mu$, $\nu$ and $g^{\mu\nu}$ refer to the Rindler coordinates ($\eta$, $\rho$, $x_\perp$) and the metrices in (\ref{astew}). 

We define $(\pi, \pi^*)$ as the canonical momenta for ($\phi^*$, $\phi$), 
and ($\bar{\pi}$, $\bar{\pi}^*$) as those with lower indices:
\begin{align}\label{fwev} 
(\pi, \pi^*) 
&\equiv (\frac{\partial {\cal L}}{\partial(\partial_\eta \phi^*)},\frac{\partial {\cal L}}{\partial(\partial_\eta \phi)}) 
=(g^{\eta\eta} \partial_\eta \phi, g^{\eta\eta} \partial_\eta \phi^*), \\*
(\bar{\pi},\bar{\pi}^*) &\equiv (g_{\eta\eta}\pi,g_{\eta\eta}\pi^*)
=(\partial_\eta \phi, \partial_\eta \phi^*). \nonumber 
\end{align} 
Corresponding to $\phi=\frac{1}{\sqrt{2}}(\phi_1+i \phi_2)$, 
$\pi_{1,2}$ and $\bar{\pi}_{1,2}$ are defined as 
\begin{align}\label{wwqp} 
\displaystyle \pi_{1,2} \equiv g^{\eta\eta}\partial_\eta \phi_{1,2}, \quad
\bar{\pi}_{1,2} \equiv g_{\eta\eta}\pi_{1,2}=\partial_\eta \phi_{1,2}.
\end{align}

With these, the Hamiltonian density associated with the Lagrangian density (\ref{sfdkdf}) is given as
\begin{align}\label{sfsdfd} 
{\cal H} 
&= \pi \,\partial_\eta \phi + \pi^* \,\partial_\eta \phi^* - {\cal L}
\nonumber \\*
&= \pi^* \bar{\pi} - g^{ij} \partial_i \phi^* \partial_j \phi +m^2 \phi^* \phi, 
\end{align}
where $i,j=\rho,x_\perp$. 
Then, we can write the Hamiltonian in the grand canonical ensemble, ${\cal H} - \mu_R \, q$, as
\begin{align}\label{powejk} 
{\cal H} - \mu_R \, q=& \, \frac{1}{2}
(g^{\eta\eta}(\bar{\pi}_1 \bar{\pi}_1+\bar{\pi}_2 \bar{\pi}_2) 
-g^{ij}(\partial _i \phi_1\partial _j \phi_1+\partial _i \phi_2\partial _j \phi_2) 
+ m^2 (\phi_1^2+\phi_2^2 )
)\nonumber\\
& -\mu_R (\pi_2 \phi_1-\pi_1 \phi_2),
\end{align} 
where $\mu_R$ is the chemical potential 
and $q \equiv -i \, g^{\eta\eta}(\phi\,\partial_\eta\phi^*-\phi^*\,\partial_\eta\phi)$ is the particle density. 
\newline

With (\ref{powejk}), we can write the probability amplitude as
\begin{align}\label{vdhtuip}
Z 
= & 
\int \! {\cal D} \bar{\pi}{\cal D} \bar{\pi}^* {\cal D} \phi \exp 
\big[ i \,\int_{\rm I} \! d^4x \, \gamma_R \, 
( g^{\eta\eta}\bar{\pi} \partial_\eta \phi + g^{\eta\eta}\bar{\pi}^* \partial_\eta \phi^* - ({\cal H} - \mu_R \, q)) \big]
\nonumber\\
= & \,\,
{\cal C}_R \int \! {{\cal D} \phi_1}{{\cal D} \phi_2}
\exp \big[ 
\frac{i}{2} \int_{\rm I} \! d^{4}x \, \gamma_R \, ( g^{\eta\eta}(\partial_\eta\phi_1+\mu_R \,\phi_2)^2 + (\partial_i \phi_1)^2 
\nonumber \\*
& \qquad \qquad \qquad \qquad \qquad \,\,
+ g^{\eta\eta}(\partial_\eta\phi_2+\mu_R \,\phi_1)^2 + (\partial_i \phi_2)^2 - m^2 ({\phi_1}^2+{\phi_2}^2))\big], 
\end{align}
where ${\rm I}$ means the whole Rindler wedge I, 
$\gamma_R \equiv \sqrt{-\det g_{\mu\nu}}$ 
and we have performed the following redefinitions for the canonical momenta:
\begin{subequations}
\begin{align}
& \bar{\pi}_1 - (\partial_\eta \phi_1 - \mu_R \,\phi_2) \rightarrow \bar{\pi}_1,\\*
& \bar{\pi}_2 - (\partial_\eta \phi_2 + \mu_R \,\phi_1) \rightarrow \bar{\pi}_2.
\end{align}
\end{subequations}
${\cal C}_R $ is given as
\begin{align} 
{\cal C}_R \equiv \int \! {\cal D} \pi_1 {\cal D} \pi_2 \exp \big[ i \int_{\Omega_{\rm I}} \! d^4x \, \gamma_R \, g_{\eta\eta} ( (\pi_1)^2+(\pi_2)^2 ) \big]. 
\end{align}
Since ${\cal C}_R$ is some number irrelevant of $\mu_R$ and we finally take the derivative with regard to $\mu_R$ to get the e.v. of the particle density, 
we can ignore ${\cal C}_R$ in our analysis in what follows. 
\newline

We perform Euclideanization:
\begin{align}\label{svpev}
\eta \to -i \tau. 
\end{align}
At this time, 
\begin{subequations}\label{vowve}
\begin{align}
& (\pi_\alpha,\bar{\pi}_\alpha) \to i(\pi_\alpha,\bar{\pi}_\alpha),\\* 
& q \to iq,\\*
\label{vowve_c}
& ds^2 \to -\rho^2 d\tau^2-d \rho^2,\quad 0 \le \tau < 2 \pi,\\*
& g_{\eta\eta} \to g_{\tau\tau},\quad g_{\tau\tau}=g_{\eta\eta},
\end{align}
\end{subequations}
where there is no changes in the contents between $g_{\tau\tau}$ and $g_{\eta\eta}$ except for the notations. 
Here, the Euclidean temperature is identical with the Unruh temperature $T_U$ in (\ref{lqvue}). 
There is a problem that the temperature depends on the space. As for this, it would be considered that 
$\rho$ is always fixed to some value corresponding to the fact particles are performing an uniformly accelerated motion.

Under the Euclideanization (\ref{svpev}) with (\ref{vowve}), $Z$ in (\ref{vdhtuip}) can be rewritten as
\begin{align}\label{vdsmh}
Z= 
\int \! {\cal D} \phi_1 {\cal D} \phi_2 
\exp \Big[ \! -\frac{1}{2} \int \! d^3 x \! \int_0^{\beta_R} \! d\tau \rho \,\gamma_R \, 
\left( \! \begin{array}{cc} \phi_1 & \phi_2 \end{array} \! \right) \! 
\left( 
\! \begin{array}{cc} 
\hat{G}_R+M_R^2 & -2ig^{\tau\tau}\mu_R \, \partial_\eta \\ 
2ig^{\tau\tau}\mu_R \,\partial_\eta & \hat{G}_R+M_R^2 
\end{array} \! 
\right) \!
\left( \! \begin{array}{c} \phi_1 \\ \phi_2 \end{array} \! \right) 
\Big],
\end{align}
where 
$\beta_R \equiv 2\pi$,
$ \hat{G}_R \equiv \, -g^{\tau\tau} \partial_\tau \partial_\tau + \gamma_R^{-1} g^{ij} \partial_i (\gamma_R\partial_j)$ and 
$M_R^2 \equiv m^2 - g^{\tau\tau}\mu_R^2$.

\subsection{Description for BEC and upper limit of chemical potential}
\label{bpwpew} 

In order to express the fields in the superconductor state, 
we separately rewrite $\phi_{1,2}$ as the e.v. part and the original $\phi_{1,2}$ like the following \cite{Kapusta:2006pm}:
\begin{subequations}\label{jhfk}
\begin{align} 
\phi_1 \, \equiv& \, \sqrt{2} \, \alpha \, \cos \Theta + \tilde{\phi}_1, \label{BEC_representation1}\\[1mm]
\phi_2 \, \equiv& \, \sqrt{2} \, \alpha \, \sin \Theta + \tilde{\phi}_2. \label{BEC_representation2}
\end{align}
\end{subequations}
where $\alpha$ and $\Theta$ mean the absolute value and the phase of e.v., and 
\begin{align}
\alpha 
\left\{
\begin{array}{l}
= 0 \quad \textrm{corresponds to the normal state,} \\[1mm]
\not= 0 \quad \textrm{corresponds to the BEC state.} \\
\end{array}
\right.
\end{align}

Rewriting $Z$ in (\ref{vdsmh}) employing the expressions in (\ref{jhfk}), we can obtain the following $Z$:
\begin{align}\label{bjskf}
Z =& 
\exp[-\alpha^2\beta_R V_\perp \int d\rho \, \rho \, \gamma_R \, M_R^2]
\int\! {\cal D} \tilde{\phi}_1 {\cal D} \tilde{\phi}_2 \exp \bigg[ 
\nonumber\\*
&
- \frac{1}{2} \int_0^\infty \! d \rho \, \rho \int_{-\infty}^\infty \! d^2 x_\perp \! \int_0^{\beta_R} \! d\tau 
\, \gamma_R \left( \! 
\begin{array}{cc} \tilde{\phi}_1 & \tilde{\phi}_2 \end{array} \! \right) \! 
\left( \! \begin{array}{cc} \hat{G}_R+M_R^2 & 2ig^{\tau\tau}\mu_R \,\partial_\eta \\ -2ig^{\tau\tau}\mu_R \,\partial_\eta & \hat{G}_R+M_R^2 \end{array} \! \right) \!
\left( \! \begin{array}{c} \tilde{\phi}_1 \\ \tilde{\phi}_2 \end{array} \! \right) 
\bigg],
\end{align}
where $V_\perp \equiv \int dx_\perp^2 = \int dk_\perp^2$.
\newline

Now, let us look at the contribution from the zero mode in the pass-integral part in (\ref{bjskf}), 
which we can write as
\begin{align}\label{alverl}
\int\! {\cal D} \tilde{\phi}_1 {\cal D} \tilde{\phi}_2 \exp [
- \frac{ M_R^2 }{2} (\tilde{\phi}_1^2 + \tilde{\phi}_2^2)
]^{V_I}, 
\end{align}
where the configurations generated by $\int {\cal D} \tilde{\phi}'_1 {\cal D} \tilde{\phi}'_2$ 
are only the constant configurations irrelevant of the coordinates, 
and $V_I$ means $\int_0^\infty \! d \rho \int_{-\infty}^\infty \! d^2 x_\perp \! \int_0^{\beta_R} \! d\tau \, \gamma_R $.
Then we can see that (\ref{bound_of_mu}) can converge when $M_R^2$ is positive, 
whereas diverge when $M_R^2$ is negative or zero. 
Therefore, we can write as
\begin{align}\label{bound_of_mu}
\textrm{$Z$ in (\ref{bjskf})} 
\left\{
\begin{array}{l}
\textrm{converges for $M_R^2>0$}, \\[1mm] 
\textrm{diverges for $M_R^2\le 0$}.
\end{array}
\right.
\end{align}
From (\ref{bound_of_mu}), we can see that there is the upper bound for the value that the chemical potential can take as
\begin{align} \label{owvu}
\mu_R^c = m/\sqrt{g^{\tau\tau}}=m/a.
\end{align}
Here, the one above is evaluated at $\xi=1$, 
therefore, we can see from (\ref{astew}) that the critical Unruh temperature obtained 
from the following analysis with (\ref{owvu}) is associated 
with the one having $\zeta$ as its proper time and performing the uniformly 
accelerated motion $a$.
\newline

Here, we explain how the BEC is formed in our system. When decreasing the Unruh temperature 
from some high temperatures (therefore, $\alpha$ is $0$) keeping the e.v. of the particle density 
constant, it turns out that the chemical potential should rise (see (\ref{density01})). However, as 
in (\ref{owvu}), there is the upper limit for the value the chemical potential can take. Therefore, 
finally $\alpha$ should start to have some finite value to keep the e.v. of the particle density constant. 
Like this, at some lower Unruh temperature, $\alpha$ becomes finite and BEC is formed. 
(For more description on this issue, see Appendix.\ref{dcmqt}.)

\subsection{Effective potential\,(1)} 
\label{beytw} 

We can diagonalize the shoulder in $Z$ in (\ref{bjskf}) as
\begin{align} \label{wvwsv}
Z =& 
\exp[-\alpha^2\beta_R V_\perp \int_0^\infty d\rho  \, \rho \,  \gamma_R M_R^2]
\int\! {\cal D} \tilde{\phi}'_1 {\cal D} \tilde{\phi}'_2 \exp 
\bigg[ 
\nonumber\\*
&
- \frac{1}{4} \int_0^\infty \! d\rho \, \rho\int_{-\infty}^\infty \! d^2 x_\perp \! \int_0^{\beta_R} \! d\tau \, \gamma_R
\left( \! \begin{array}{cc} \tilde{\varphi}'{}^* & \tilde{\varphi}' \end{array} \! \right) \!
\left( \! \begin{array}{cc} 
\hat{G}_{R+} + M_R^2 & 0 \\ 0 & \hat{G}_{R-} + M_R^2
\end{array} \! \right)
\!
\left( \! \begin{array}{c} \tilde{\varphi}' \\ \tilde{\varphi}'{}^* \end{array} \! \right)
\bigg], 
\end{align}
where 
$\hat{G}_{R\pm} \equiv \hat{G}_R \pm 2g^{\eta\eta}\mu_R$ and
$
\frac{1}{\sqrt{2}} \left( \! \begin{array}{c} \tilde{\varphi}' \\ \tilde{\varphi}'{}^* \end{array} \! \right) 
\equiv U^{-1} \left( \! \begin{array}{c} \tilde{\phi}_1 \\ \tilde{\phi}_2 \end{array} \! \right)
$. 
$U$ is given as $\frac{1}{\sqrt{2}} \left( \begin{array}{cc} i & -i \\ 1 & 1 \end{array} \right)$. 
The difference arisen in the path-integral measure by the transformation $U$ is just some constant, which we can ignore. 

We express $\phi_\alpha$ $(\alpha=1,2)$ by the plain wave expansion for the $x_\perp$-directions remaining the $\rho$-direction as
\begin{align}\label{ovfvsd}
\tilde{\phi}_{\alpha}(\rho,\eta,x^\perp) 
= \frac{1}{(2\pi)^2\beta_R} \sum_{n=-\infty}^\infty \int dk_\perp^2 \tilde{\varphi}_{\alpha,n} (\rho,k_\perp) e^{-i(\omega_n \eta\,+\,k_\perp x^\perp)}.
\end{align}
Then $\hat{G}_0 \equiv \frac{1}{2}(\hat{G}_{R+} + \hat{G}_{R-}+2 M_R^2)$ in $Z$ in (\ref{wvwsv}) is given as\footnote{
$\hat{G}_0$ appears in (\ref{wvwsv}) as follows:
\begin{align}
\left( \! \begin{array}{cc} \tilde{\varphi}'{}^* & \tilde{\varphi}' \end{array} \! \right) \!
\left( \! \begin{array}{cc} 
\hat{G}_{R+} + M_R^2 & 0 \\ 0 & \hat{G}_{R-} + M_R^2
\end{array} \! \right)
\!
\left( \! \begin{array}{c} \tilde{\varphi}' \\ \tilde{\varphi}'{}^* \end{array} \! \right)
=(\hat{G}_{R+} + \hat{G}_{R-}+2 M_R^2) |\tilde{\varphi}'|^2. 
\end{align}
}
\begin{align}\label{lsdgv}
& 
\hat{G}_0 \to 
G_0 = g^{\tau\tau} \omega_n^2 + \gamma_R^{-1} g^{\rho\rho} \partial_\rho (\gamma_R\partial_\rho) + g^{\perp\perp} k_\perp^2+M_R^2
\equiv G_R+M_R^2
\end{align}
with $(\partial_\eta,\partial_\perp) \to -i(\omega_n, k_\perp)$; 
$ \gamma_R^{-1} g^{\rho\rho} \partial_\rho (\gamma_R \partial_\rho)$ is still operator. 
As a result, $Z$ in (\ref{wvwsv}) can be given as
\begin{align}\label{osmytp}
Z =
& 
\exp[-\alpha^2\beta_R V_\perp \int_0^\infty d\rho \, \rho \,\gamma_R M_R^2]
\int\! {\cal D} \tilde{\phi}'_{1,n} {\cal D} \tilde{\phi}'_{2,n} \exp \bigg[ 
\nonumber\\*
&
- \frac{1}{2\beta_R} \sum_{n=-\infty}^\infty \int_0^\infty \! d\rho \, \rho \int_{-\infty}^\infty \! d^2 k_\perp \, \gamma_R
(G_R+M_R^2)
( (\tilde{\phi}'_{1,n}(\rho,k_\perp))^2 + (\tilde{\phi}'_{2,n}(\rho,k_\perp))^2 )
\bigg],
\end{align}
where the general formula: $\beta^{-1}\int_0^\beta d\tau e^{i\tau(\omega_m-\omega_n)}=\delta_{mn}$ 
has been used. 
Then, performing the functional integral for $\tilde{\phi}'_{\alpha,n}$, 
we can get the following $Z$: 
\begin{align}\label{oregev}
Z 
=
\exp[-\alpha^2\beta_R V_\perp \int d\rho \,\rho\,\gamma_R M_R^2]
\prod_{n=-\infty}^\infty
\prod_\rho
\Det (\frac{\pi\,\rho\Delta \rho \, \gamma_R}{2\beta_R} (G_R+M_R^2))^{-1}, 
\end{align}
where from (\ref{osmytp}) to (\ref{oregev}), 
we have rewritten the integral $\int d\rho$ as $\sum_\rho \Delta \rho$ ($\rho$ takes all the real numbers from $0$ to $\infty$), 
and $\Det$ is the one with regard to the $k_\perp$-space for each $\rho$, 
where we have written the reason for why we separate by each $\rho$ in Sec.\ref{ksdvb}. 
~\newline

Defining the free energy $F_R$ as $Z=\exp (-\beta_R F_R )$, 
\begin{align}\label{nwwes} 
F_R 
=& \, 
\alpha^2 V_\perp \int_0^\infty d\rho \rho\,\gamma_R M_R^2
+
\frac{1}{\beta_R}
\sum_{n=-\infty}^\infty \sum_\rho \rho\,{\rm Tr}\, {\rm Log} (G_R+M_R^2),
\end{align} 
where we have ignored ${\cal N} \, {\rm Log}(\frac{\pi\Delta \rho \, \gamma_R}{2\beta_R})$ (where ${\cal N}={\rm Dim}(n) \times {\rm Dim}(\rho) \times {\rm Dim}(k_\perp)$) as 
it is irrelevant of $\mu_R$. We can express $F_R$ in (\ref{nwwes}) as
\begin{align}\label{nytes} 
F_R
=& \, 
\alpha^2 V_\perp \int_0^\infty d\rho \,\rho\,\gamma_R M_R ^2
+
\frac{V_\perp}{\beta_R}
\sum_{n=-\infty}^\infty \sum_\rho \rho\,\gamma_R \int \frac{dk_\perp^2}{(2\pi)^2}
\int_0^{M_R^2}d\Delta^2 \widetilde{D}_{\Delta^2}(\rho,n, k_\perp), 
\end{align} 
where
\begin{align}\label{vdsv}
\widetilde{D}_{\Delta^2}(\rho, k_\eta,k_\perp) 
&\equiv 
\left(G_R + \Delta^2 \right)^{-1} 
= 
(
g^{\tau\tau} \omega_n^2 + \gamma_R^{-1} g^{\rho\rho} \partial_\rho (\gamma_R \partial_\rho) + g^{\perp\perp} k_\perp^2 + \Delta^2)^{-1},
\end{align} 
where 
\begin{align}\label{svergmd}
\omega_n=n.
\end{align} 
This is because $\tau$ is periodic with the period $2\pi$ as can be seen (\ref{vowve_c}), 
therefore  $\omega_n$ is given as $\omega_n = 2\pi n /\beta_R$  $(\beta_R=2\pi)$. 
(Here, the Euclidean temperature is identical with the Unruh temperature (\ref{lqvue}))

In (\ref{nytes}), we have ignored ${\rm Log} \, G_R$ as it is irrelevant of $\mu_R$\footnote{
${\rm Log} \, G_R$ appears in (\ref{nytes}) as follows:
\begin{align}
{\rm Log} (G_R+M_R^2)=\int_0^{M_R^2}d\Delta^2 \left(G_R + \Delta^2 \right)^{-1} +{\rm Log} G_R.
\end{align} 
}. 
Further, $\widetilde{D}_{\Delta^2}(\rho, k_\eta,k_\perp)$ in (\ref{nytes}) includes some operator, 
but for now we suppose that it is some numbers. 

Considering that the (\ref{nytes}) divided by $\rho$ is the effective potential for the particles 
performing a uniformly accelerated motion determined by $\rho$,
we consider to express $F_R/V_\perp$ for each $\rho$ as $\Gamma_R^{(\rho)}$, then,
\begin{align}\label{W_Dpm}
\Gamma_R^{(\rho)} &= 
\alpha^2 \gamma_R M_R ^2
+\frac{1}{\beta_R } 
\sum_{n=-\infty}^\infty
\gamma_R \! 
\int_{-\infty}^\infty \! 
\frac{dk^2}{(2\pi)^2} 
\int^{M_R^2}_0 \!\!\! d\Delta^2 \, 
\widetilde{D}_{\Delta^2}(\omega_n,\rho,k).
\end{align} 

\subsection{$\widetilde{D}_{\Delta^2}(\omega_n,\rho,k)$}
\label{pwmnv} 

$\widetilde{D}_{\pm,\Delta^2}(\rho, \omega_n,k_\perp)$ in (\ref{vdsv}) is given as some operator. 
In this subsection, we obtain its expression as numbers.
For this purpose, we define $\widetilde{D}_{\pm,\Delta^2}(\rho, \omega_n,k_\perp)$ as
\begin{align}\label{kldasn}
D_{\Delta^2}(x-x')
=\int \! \frac{d^3k}{(2\pi)^3} \, \widetilde{D}_{\Delta^2}(\rho-\rho',k_\eta,k_\perp) \, e^{i\,(\omega_n \, (\eta-\eta') - k\!_\perp (x\!_\perp-x\!_\perp'))}.
\end{align}
$D_{\Delta^2}(x-x')$ should satisfy the following identity: 
\begin{align}\label{Identity_of_Dpm}
& (
g^{\tau\tau} \omega_n^2 + \gamma_R^{-1} g^{\rho\rho} \partial_\rho (\gamma_R \partial_\rho) + g^{\perp\perp} k_\perp^2 + \Delta^2
) D_{\Delta^2}(x-x')
=\gamma_R^{-1} \delta^4(x-x'),
\end{align}
where the operator in l.h.s. in (\ref{Identity_of_Dpm}) is taken from (\ref{nytes}).
From (\ref{Identity_of_Dpm}), we can obtain the relation that $\widetilde{D}_{\pm,\Delta^2}$ should satisfy as
\begin{align}\label{GreenFuncEOM2}
& \big(\,\hat{{\cal F}} \green{-} \omega_n^2\,\big) \,\widetilde{D}_{\pm,\Delta^2}(\rho-\rho',k_\eta,k_\perp) 
= -\rho \, \delta(\rho-\rho'), \\*
& \hat{{\cal F}} \equiv \rho^2 \partial_\rho^2+ \rho \, \partial_\rho - \rho^2\kappa^2, \quad
\kappa^2 \equiv (k_\perp^2+\Delta^2).
\nonumber
\end{align}
Based on (\ref{GreenFuncEOM2}), we obtain $\widetilde{D}_{\Delta^2}(\rho-\rho',k_\eta,k_\perp)$ in what follows.
\newline

To obtain $\widetilde{D}_{\Delta^2}(\rho-\rho',k_\eta,k_\perp) $, 
we focus on the fact that 
$\hat{{\cal F}}$ is the operator of the following eigenvalue equation with eigenvalue $(i\lambda)^2$:
\begin{align}\label{lsdvn}
&\hat{{\cal F}} \,\Theta_\lambda(\rho\kappa) = (i \lambda)^2 \, \Theta_\lambda(\rho\kappa), \\*
&\Theta_\lambda(\rho\kappa)\equiv C_\lambda K_{i \lambda}(\rho\kappa),\quad C_\lambda \equiv \frac{1}{\pi}\sqrt{2\lambda \sinh(\pi \lambda)}, \nonumber
\end{align}
where $K_\alpha(x)$ is the second kind modified Bessel function, 
and $\Theta_\lambda(\rho,k)$ satisfies the following normalized orthogonal relation\footnote{
It is written in \cite{Fulling:1972md,Terashima:1999xp} as
$\int_0^\infty \frac{dx}{x} \, K_\alpha(x) \, K_{\beta}(x)=\frac{\pi^2 \delta(\alpha-\beta)}{2\beta \sinh \pi \beta}$, 
which we derive in Appendix.\ref{dgwfdn}.
Writing $x$ with $ay$, l.h.s. can be changed to $\int_0^\infty \frac{dy}{y} \, K_\alpha(ay) \, K_{\beta}(ay)$ 
without changing the r.h.s. 
}:
\begin{align}\label{sdwek}
\int_0^\infty \frac{d\rho}{\rho} \, \Theta_{\lambda'}(\rho\kappa) \, \Theta_\lambda(\rho\kappa)=\delta(\lambda'-\lambda). 
\end{align}
$\lambda$ is real number, therefore $\Theta_\lambda(\rho\kappa)$ can form a set of infinite dimensional orthogonal system. 
Then, by taking $\Theta_\lambda(\rho\kappa)$ as a set of the orthogonal bases, 
let us formally write $\widetilde{D}_{\pm,\Delta^2}(\rho-\rho',k_\eta,k_\perp)$ in the expanding form as
\begin{align}\label{Theta_expansion_of_D}
\widetilde{D}_{\Delta^2}(\rho-\rho',k_\eta,k_\perp)
=\int_0^\infty \! d\lambda \, f_{\lambda} \, \Theta_\lambda(\rho\kappa), 
\end{align}
where $f_{\lambda,\pm}$ are the coefficients of each independent direction, $\Theta_\lambda(\rho\kappa)$, specified by $\lambda$, 
which are to be obtained in what follows.
\newline

To obtain $f_{\lambda}(\omega,\rho')$, we consider two quantities: 
$\langle \widetilde{D}_{\Delta^2} | \hat{{\cal F}} | \Theta_\lambda(\rho\kappa) \rangle$ 
and $\langle \Theta_\lambda(\rho\kappa) | \hat{{\cal F}} | \widetilde{D}_{\Delta^2} \rangle$, 
which should be equivalent each other, however each of these expressions can be given as
\begin{subequations}
\begin{align} 
\label{DFTheta}
&\int_0^\infty \frac{d\rho}{\rho} \, \widetilde{D}_{\Delta^2} \, \hat{{\cal F}} \, \Theta_\lambda(\rho\kappa)
= -\lambda^2 \int_0^\infty \frac{d\rho}{\rho} \, \widetilde{D}_{\Delta^2} \, \Theta_\lambda(\rho\kappa), \\*
\label{ThetaFD}
&\int_0^\infty \frac{d\rho}{\rho} \, \Theta_\lambda(\rho\kappa) \, \hat{{\cal F}} \widetilde{D}_{\Delta^2}
= \green{\omega_n^2} \int_0^\infty \frac{d\rho}{\rho} \, \Theta_\lambda(\rho\kappa) \widetilde{D}_{\Delta^2}
- \Theta_\lambda(\rho'\kappa),
\end{align}
\end{subequations} 
where we have used (\ref{lsdvn}) and (\ref{GreenFuncEOM2}). 
From the condition $(\ref{DFTheta})-(\ref{ThetaFD})=0$ where\footnote{
(\ref{wuef}) is calculated as follows:
\begin{align} 
(\ref{DFTheta})-(\ref{ThetaFD})
&=\big(\! -\lambda^2+\omega_n^2 \big) \int \frac{d\rho}{\rho} \widetilde{D}_{\Delta^2} \, \Theta_\lambda(\rho\kappa) 
+ \Theta_\lambda(\rho'\kappa)\nonumber \\*
&= 
\big(\! -\lambda^2+\omega_n^2 \big) 
\int \frac{d\rho}{\rho} \cdot \int \!d\lambda' \, f_{\lambda'} \Theta_{\lambda'}(\rho\kappa) \cdot \Theta_\lambda(\rho\kappa) 
+ \Theta_\lambda(\rho'\kappa)\nonumber \\*
&= \big(\! -\lambda^2+\omega_n^2 \big)
\int \!d\lambda'
\cdot 
\int \frac{d\rho}{\rho}\, \Theta_{\lambda'}(\rho\kappa)\Theta_\lambda(\rho\kappa)
\cdot 
f_{\lambda'}
+ \Theta_\lambda(\rho',k)\nonumber \\*
&= \big(\! -\lambda^2+\omega_n^2 \big)
\int \!d\lambda'
\cdot 
\delta(\lambda'-\lambda)
\cdot 
f_{\lambda'} 
+ \Theta_\lambda(\rho'\kappa)\nonumber \\*
&
= \big(\! -\lambda^2+\omega_n^2 \big) \, f_{\lambda}
+ \Theta_\lambda(\rho'\kappa).
\end{align}}
\begin{align} \label{wuef}
(\ref{DFTheta})-(\ref{ThetaFD}) = \big(\! -\lambda^2\green{-}\omega_n^2 \big) \, f_{\lambda} + \Theta_\lambda(\rho'\kappa),
\end{align}
$f_{\lambda,\pm}(\rho-\rho',k_\eta,k_\perp)$ can be determined as
\begin{align}\label{f_result}
f_{\lambda}=\frac{\Theta_\lambda(\rho'\kappa)}{\lambda^2\green{+}\omega_n^2}. 
\end{align}
Using (\ref{f_result}) in (\ref{Theta_expansion_of_D}), we can write $\widetilde{D}_{\Delta^2}$ as
\begin{align}\label{evuo}
\widetilde{D}_{\Delta^2}
(\rho-\rho',k_\eta,k_\perp)
=\int_0^\infty \! d\lambda \, \frac{\Theta_\lambda(\rho'\kappa)\Theta_\lambda(\rho\kappa)}{\lambda^2\green{+}\omega_n^2}
=\int_0^\infty \! d\lambda \, \frac{C_\lambda^2 K_{i \lambda}(\rho'\kappa)K_{i \lambda}(\rho\kappa)}{\lambda^2\green{+}\omega_n^2}.
\end{align} 
This expression does not include operators, which is some numbers in principle.

\subsection{Effective potential\,(2)}
\label{sdjkvbsd}

In the previous subsection we have obtained some concrete expression of $\widetilde{D}_{\Delta^2}(\rho-\rho',k_\eta,k_\perp)$ as in (\ref{evuo}).
Using it, we can write $\Gamma^{(\rho)}_R$ in (\ref{W_Dpm}) as\footnote{
We have used
$\displaystyle \sum_{n=-\infty}^\infty \frac{1}{\lambda^2+n^2}=\frac{\pi }{\lambda}\coth (\pi \lambda)$ and 
$\displaystyle \frac{1}{\beta_R}\left(\frac{\sqrt{2\lambda}}{\pi}\right)^2\int_{-\infty}^\infty \frac{dk^2}{(2\pi)^2}
=\frac{\lambda}{2\pi^4}\int_0^\infty dk k$.}.
\begin{align}\label{W_Dpm2}
\Gamma^{(\rho)}_R =
\alpha^2 \gamma_R M_R^2
+\frac{\gamma_R}{2\pi^3} 
\int_0^\infty \! d\lambda \, \cosh(\pi \lambda) \int_0^\infty \! dk k\, \Psi,
\end{align}
where 
\begin{subequations} 
\begin{align}\label{Psi_def}
\Psi &\equiv \int^{M_R^2}_0 \!\! d\Delta^2 \, K_{i\lambda}^2(\kappa\rho), \quad (\kappa^2 \equiv k_\perp^2+\Delta^2, \quad M_R^2 \equiv m^2 - g^{\eta\eta}\mu_R^2).
\end{align}
\end{subequations} 

Considering to perform the derivative with regard to $\mu_R$, 
we pick up the $\mu_R$-dependent parts in $\Psi$ by expanding it around $\mu_R=0$\footnote{
In this section we have performed the expansion around $\mu_R=0$ as in (\ref{dcsl}). 
However, if we expanded around $\mu_R=\mu_R^c$, we can obtain the following effective potential:
\begin{align}\label{bsdpw}
\Gamma^{(\rho)}_R= 
\alpha^2 a_c^{-1}M^2
+\frac{a_c}{\pi^3} 
\int_0^\infty d \lambda \cosh \pi \lambda 
\int_0^\infty dk k K_{i\lambda}^2 (\frac{m}{a} \mu_R -\frac{m^2}{a^2}). 
\end{align}
Therefore, there is no difference in the e.v. of the particle density we can obtain finally, .
} 
as
\begin{align}\label{dcsl}
\Psi =& \, \Psi\Big|_{\mu_R=0} 
+ \frac{\partial \Psi}{\partial \mu_R} \Big|_{\mu_R=0} \,\mu_R 
+ \frac{1}{2}\frac{\partial ^2 \Psi}{\partial \mu_R^2} \Big|_{\mu_R=0} \,\mu_R^2 
+ {\cal O}(\mu_R^4)
\nonumber\\*
=& \Psi_0- a^2 K_{i\lambda}^2(\kappa\rho) \Big|_{\Delta^2=M^2} \, \mu_R^2 + {\cal O}(\mu_R^4)
\end{align}
where\footnote{
We have calculated (\ref{cwoe2}) and (\ref{cwoe3}) using the following formula: 
\begin{align}
\frac{\partial}{\partial \beta} \int_\alpha^\beta dx f(x) 
= \lim_{\Delta \beta \to 0}\frac
{1}{\Delta \beta}(\int_\alpha^{\beta+\Delta\beta} dx f(x) - \int_\alpha^\beta dx f(x))
= \lim_{\Delta \beta \to 0}\frac
{1}{\Delta \beta}( f(\beta+\Delta \beta)\Delta \beta+ {\cal O}(\Delta \beta^2))
.\end{align}
}, 
\begin{subequations}
\begin{align}
\label{cwoe1}
\Psi_0 &\equiv \int_0^{m^2} \!\! d \Delta^2 \, K_{i\lambda}^2(\kappa\rho) ,\\*
\label{cwoe2}
\frac{\partial \Psi}{\partial \mu_R} \bigg|_{\mu_R=0} 
&= 
\frac{\partial M_R^2}{\partial \mu_R} \frac{\partial \Psi}{\partial M_R^2} \bigg|_{\mu_R=0} = 0,\\*
\label{cwoe3}
\frac{1}{2}\frac{\partial ^2 \Psi}{\partial \mu_R^2} \bigg|_{\mu_R=0}
&=
\frac{1}{2}\frac{\partial }{\partial \mu_R} \bigg(\frac{\partial M_R^2}{\partial \mu_R} \frac{\partial \Psi}{\partial M_R^2} \,\bigg)\bigg|_{\mu_R=0}
=
-a^2 K_{i\lambda}^2(\kappa\rho) \Big|_{\Delta^2=m^2}.
\end{align}
\end{subequations} 
With (\ref{dcsl}), we can write $\Gamma^{(\rho)}_R$ in (\ref{W_Dpm2}) as
\begin{align}\label{density2}
\Gamma^{(\rho)}_R = 
\alpha^2\gamma_R M_R^2
+
\frac{\gamma_R}{2\pi^3} \int_0^\infty d\lambda \, \cosh(\pi \lambda) 
\int_0^\infty \! dk k\, 
\Big(
\Psi_0 
-a^2\mu_R^2 \, K_{i\lambda}^2(\kappa\rho)\Big|_{\Delta^2=m^2} \Big).
\end{align}
~\newline

We are going to finally assign the critical value $\mu_R^c=m/a_c$ in (\ref{bound_of_mu}) to $\mu_R$ in (\ref{dcsl}), 
then take the leading order in the expansion of $\mu_R^c$.
For this, now we use two symbols, $\mu_R$ and $\mu_R'$:
\begin{itemize}
\item $\mu_R$: chemical potential on which the derivative regarding $\mu_R$ can act; finally the value $m/a_c$ is assigned, 
\item $\mu_R'$: just a symbol for the value $m/a_R^c$, on which the derivative regarding $\mu_R$ does not act on. 
\end{itemize}

Then, $\Psi_0$ and $a^2\mu_R^2 \, K_{i\lambda}^2(\kappa\rho)\Big|_{\Delta^2=m^2}$ can be expanded with regard to $\mu_R'$ as 
\begin{subequations}
\label{Expanded_Psi}
\begin{align}
\Psi_0 =& \,\, 
a_c^2 \, K_{i \lambda}^{2}(k/a_c)\,{\mu_R'}^2
+{\cal O}(\mu_R'{}^{4}), \\*
a_c^2\mu_R^2 \, K_{i\lambda}^2(\kappa\rho)\Big|_{\Delta^2=m^2} =& 
a_c^2\mu_R^2\,(K_{i \lambda}^2(k/a_c)+{\cal O}(\mu_R'{}^2)). 
\end{align}
\end{subequations}
Using (\ref{Expanded_Psi}), 
we can obtain $\Gamma^{(\rho)}_R$ at the critical moment given by (\ref{dcsl}) as 
\begin{align}\label{bwveovi}
\Gamma^{(\rho)}_R= \alpha^2a_c^{-1}M_R^2+\frac{a_c^3}{4\pi^2} (\mu_R'{}^2-\mu_R^2)\int^\infty_0 d\lambda \lambda \coth (\pi \lambda),
\end{align}
where we have used $\displaystyle \int^\infty_0 dkk K_{i\lambda}^2(k/a_c)$ is given as $\frac{a_c^2\pi \lambda}{2\sinh \pi \lambda}$.
~\newline

Let us evaluate the $\lambda$-integral in (\ref{bwveovi}). 
\begin{align}\label{ncpwc}
\int^\infty_0 d\lambda \lambda \coth (\pi \lambda). 
\end{align}
First, we can see that (\ref{ncpwc}) is diverged if it is evaluated as it is. 
Therefore, we consider to do some regularize toward (\ref{ncpwc})\footnote{
The divergences also appear in other works 
for the critical acceleration for the spontaneous symmetry breaking \cite{Ohsaku:2004rv,Ebert:2006bh,Castorina:2012yg} 
and the $D=1+3$ Euclid space at finite temperature \cite{Kapusta:2006pm}. 
In these, the divergences are ignored supposing that some regularization, e.g. a mass renormalization, the UV-cutoff and so on, could work, though it is not shown explicitly. 
}. 
For this purpose, we consider to pull out some constant in $\coth (\pi \lambda)$. 
Therefore, expressing $\coth (\pi \lambda)$ as
\begin{align}\label{nslvr}
\coth (\pi \lambda)= 1+\frac{2}{e^{2 \pi \lambda }-1}, 
\end{align}
we exclude ``$1$''. 
Then, once putting the upper limit of the integral as $\Lambda$ we perform the integral, then we get as
\begin{align}\label{nwsthe}
\int^{\Lambda}_0 d\lambda \lambda (\coth (\pi \lambda)-1)
=
-\frac{1}{12} 
-\Lambda ^2
+\frac{\Lambda \log (1-e^{2 \pi \Lambda })}{\pi }+\frac{\text{Li}_2(e^{2 \pi \Lambda })}{2 \pi ^2},
\end{align}
where $\Lambda \to \infty$ finally. 
Then excluding ``$-\frac{1}{12}$'' in (\ref{nwsthe}), we take the $\Lambda$ to $\infty$. 
As a result we get as
\begin{align}\label{vdouyr}
\lim_{\Lambda \to \infty}
(-\Lambda ^2+\frac{\Lambda \log (1-e^{2 \pi \Lambda })}{\pi }+\frac{\text{Li}_2(e^{2 \pi \Lambda })}{2 \pi ^2})=\frac{1}{6}. 
\end{align}
Then, using this result, we can give the effective potential $\Gamma_R^{(\rho)}$ as
\begin{align}\label{btncew}
\Gamma_R^{(\rho)}= \alpha^2a_c^{-1}M_R^2+\frac{a_c^3}{24\pi^2}(\mu_R'{}^2-\mu_R^2). 
\end{align}
\subsection{Critical Unruh temperature for BEC}
\label{vlsdkv}

We can obtain the e.v. of the particle density according to $d_R = - \partial\Gamma_R^{(\rho)}/\partial\mu_R$ as
\begin{align}\label{density01}
d_R = 2\alpha^3 a_c \mu_R + \frac{a_c^3}{12\pi^2}\mu_R, 
\end{align} 
From (\ref{density01}), we can see that 
if we decrease the Unruh temperature from some high temperatures fixing the e.v. of the particle density to some constant,
the value of the chemical potential should rise.

Let us consider to reach the critical Unruh temperature 
by decreasing the acceleration gradually from some high accelerations where BEC is not formed. 
Then, from the explanation under (\ref{owvu}), we can obtain the critical acceleration from (\ref{density01}) as\footnote{
(\ref{result}) can be written in the MKS units as 
$
a^c = \, 2 \, \pi \, \sqrt{c^3\hbar\,3d_R/m}
\,\approx 
16.763\,\sqrt{2\pi}\,\sqrt{d_R/m} \, [ {\rm cm}/{\rm s}^2 ],
$
where $m$ is the mass of the particle comprising the gas and $d_R$ is its density. 
}
\begin{align}\label{result}
a^c = 2 \, \pi \,\sqrt{3d_R/m},
\end{align}
where $\alpha=0$ and $\mu_c = m/a^c$ have been assigned in (\ref{density01}). 
Here, note the comment given under (\ref{owvu}).

Using the relation between Unruh temperature and acceleration, $a=2\pi T_U$, 
we can obtain the critical Unruh temperature as $T_U = \sqrt{3d_R/m}$. 
This result is consistent with the critical temperature for BEC obtained from the different analytical method 
in the $D=1+3$ flat Euclid space at finite temperature \cite{Kapusta:2006pm}.

\section{Critical Unruh temperature of BEC in flat Euclid space at finite temperature} 
\label{qbele} 

In this section, we obtain the critical temperature in the $D=1+3$ flat Euclidean space at finite temperature 
from our analysis in the previous section just by exchanging the space time for the $D=1+3$ flat Euclidean space at finite temperature. 

Therefore, as the calculation way in this section is basically the same with that in the previous section, 
we in this section describe only the points in the case of the $D=1+3$ flat Euclidean space at finite temperature.

\subsection{Exchange the back ground space for Euclidean space}
\label{qqcwe} 

First, we exchange the Rindler space (\ref{astew}) for the flat $D=1+3$ Euclidean space at finite temperature, which can be done 
by {\bf 1)} putting $\rho$ to $1$, 
then {\bf 2)} Euclideanizing $\tau$-direction like (\ref{svpev}) 
then periodizing it by the arbitrary period $\beta_E$. 
The (\ref{astew}) in which these 2 manipulations are performed can be written as
\begin{align}\label{alkvd}
ds_E^2 = - d\tau^2 - d\rho^2 - dx_\perp^2, 
\end{align}
where the $\tau$-direction is periodic with the arbitrary period $\beta_E$.

\subsection{Effective potential}
\label{qcsetrb} 
Employing (\ref{alkvd}), we proceed with the analysis in previous section. 
Then, the following $\Gamma_E$ can be obtained instead of (\ref{W_Dpm}):
\begin{align}\label{xbpeo}
\Gamma_E =& \,\, 
\alpha^2M_E^2+
\frac{2}{\pi^2\beta_E} 
\int_{-\infty}^\infty \frac{dk^2\!_\perp}{(2\pi)^2} 
\int_0^\infty \! d\lambda \, \lambda \, \sinh(\pi \lambda)
\sum_{n=-\infty}^\infty \! \frac{1}{\lambda^2-\omega_n^2} \Psi_E,
\end{align}
where 
\begin{align}\label{wvire}
\omega_n\equiv \frac{2\pi}{\beta_E}n, \quad 
\Psi_E \equiv \int^{M_E^2}_0 \!\!\! d\Delta^2 \, K_{i\lambda}^2(\kappa), \quad ( \kappa^2 \equiv k_\perp^2+\Delta^2, \quad M_E^2 \equiv m^2 - \mu_E^2), 
\end{align}
and $K_\alpha(x)$ mean the second-kind Bessel function, 
and as for $\omega_n$, we use the same notation with the one in the previous subsection.

From $M_E^2$ above, we can see that 
the upper bound of the value that the chemical potential in the case of the flat $D=1+3$ Euclidean space at finite temperature 
is given as
\begin{align}\label{dsvuje}
\mu_E^c=m.
\end{align}
Expanding (\ref{xbpeo}) to the second-order of the value of the critical chemical potential in the same way we have done in Sec.\ref{sdjkvbsd}, 
we can obtain the following $\Gamma_E$:
\begin{align}\label{xbdvit}
\Gamma_E = \, 
\alpha^2M_E^2+
\frac{1}{4\pi^2} 
\int_0^\infty \! d \lambda \, \lambda \, \coth(\frac{\beta_E \lambda}{2}) (\mu_E'^2-\mu_E^2),
\end{align}
where $\mu_E'$ is the same meaning with that in Sec.\ref{sdjkvbsd}, 
but in the current case, $\mu_E'$ is given by $m$ corresponding to (\ref{dsvuje}). 

Now, we evaluate $\displaystyle \int_0^\infty \! d \lambda \, \lambda \, \coth(\beta_E \lambda/2)$ in (\ref{xbdvit}). 
If we performed the integration as it is, it would be diverged. 
Hence, we do some regularization. 
As $\coth x$ can be rewritten as $\displaystyle 1+{2}/(e^{2x}-1)$, 
we subtract the constant ``1'' as some regularization, then evaluate it as
$\displaystyle \int_0^\infty \! d \lambda \, \lambda \, ( \coth(\beta_E \lambda/2)-1)=2\pi^2/3\beta_E^2$.
Using this, we can obtain the following $\Gamma_E$:
\begin{align}\label{xvykdv}
\Gamma_E = \, 
\alpha^2M_E^2+
\frac{1}{6\beta_E^2} 
(\mu_E'^2-\mu_E^2). 
\end{align}

\subsection{Critical temperature}
\label{vwbttdc} 

From (\ref{xvykdv}), 
calculating the density according to $d_E=-\frac{\partial \Gamma_E}{\partial \mu_E}$, 
then putting $\alpha=0$ and $\mu_E^c=m$ corresponding to the critical moment, 
we can obtain the relation between the temperature and density at the critical moment as
\begin{align}\label{dccile}
T_E^c=\sqrt{3d_E/m}.
\end{align}
This can agree to the critical temperature for BEC in the $D=1+3$ flat Euclid space at finite temperature in \cite{Kapusta:2006pm}, 
however our way to obtain this result is different from \cite{Kapusta:2006pm}.

\section{BEC in the traversable wormhole space time}
\label{bsbrtw} 

\subsection{Traversable wormhole space time}
\label{bermfd} 

In this section, we consider the traversable wormhole given 
from considering two of the following space time:
\begin{eqnarray}\label{qdjes} 
ds^2 = dt^2 - \frac{dr^2}{1-v^2/r^2}-r^2(d\theta^2+\sin^2 \theta d\phi^2),
\end{eqnarray} 
then attaching the parts of $r=v$ of these \cite{Visser:1995cc}.
Therefore, the position of the throat is located at $r=v$ and the range of $r$ in one side of the wormhole space time is 
\begin{eqnarray}\label{wnyke} 
v \le r \le \infty, \quad (\textrm{the throat is located at $r=v$}). 
\end{eqnarray} 

In order that the wormhole (\ref{qdjes}) can satisfy the Einstein equation,  
the exotic matter, some matter to violate the energy condition, 
is needed \cite{Visser:1995cc} as shown in the next section, however we do not consider it in this study.

\subsection{\bf{On the fact that matter to violate the energy condition is needed}}
\label{vvbep}

(\ref{qdjes}) leads to the following Einstein tensor:
\begin{eqnarray}\label{codc} 
R_{\mu\nu}-\frac{g_{\mu\nu}}{2}R=
\left(
\begin{array}{cccc}
-\frac{v^2}{r^4} & 0 & 0 & 0 \\
0 & \frac{v^2}{r^2 \left(v^2-r^2\right)} & 0 & 0 \\
0 & 0 & \frac{v^2}{r^2} & 0 \\
0 & 0 & 0 & \frac{v^2 \sin ^2 \theta}{r^2} \\
\end{array}
\right).
\end{eqnarray}

From the result above, we can see that the energy density of our scalar field should be negative, 
which means that our scalar field is tachyonic or to break  the energy condition. 
However, since such tachyonic matter cannot exist, (\ref{qdjes}) cannot be a solution realistically. 
Therefore, (\ref{qdjes}) will not be realized. 
This study treats (\ref{qdjes}) upon knowing this. 
If we care this problem in this study, we could mention that 
the wormhole space time appears for some reason and exists for 
some time, then this study is the one for that time.

Indeed, this problem is general in the context of the traversable wormhole space time. 
For the references for the recent studies on this, see Sec.\ref{jeloiw}.

\subsection{Probability amplitude and its Euclideanization}
\label{bsawe} 

In the case that the wormhole (\ref{qdjes}) is taken as our space time, the probability amplitude corresponding to (\ref{sfsdfd}) is given as 
\begin{align}\label{vboqe}
Z 
= & \,\,
{\cal C}_w \! {{\cal D} \phi_1}{{\cal D} \phi_2}
\exp \big[ 
\frac{i}{2} \int \! d^{4}x \, \gamma_w \, ( j^{tt}(\partial_t\phi_1+\mu_w \,\phi_2)^2 + (\partial_i \phi_1)^2 
\nonumber \\
& \qquad \qquad \qquad \qquad \qquad \,\,
+ j^{tt}(\partial_t\phi_2+\mu_w \,\phi_1)^2 + (\partial_i \phi_2)^2 - m^2 ({\phi_1}^2+{\phi_2}^2))\big], 
\end{align}
where $i=r,\theta,\phi$ and $j_{\mu\nu}$ refer to the metrics in (\ref{qdjes}) and $\gamma_w \equiv \sqrt{-\det j_{\mu\nu}}$. 
${\cal C}_w$ is ignorable as well as Sec.\ref{bsow}. 

We perform the Euclideanization, 
\begin{align}\label{qvsfm}
t \to -i \tau. 
\end{align}
At this time, 
\begin{subequations}\label{qrvurs} 
\begin{align}
& (\pi_\alpha,\bar{\pi}_\alpha) \to i(\pi_\alpha,\bar{\pi}_\alpha) ,\\*
& dt^2 \to -d\tau^2,\\* 
& j_{tt} \to j_{\tau\tau} \quad (j_{tt} =j_{\tau\tau}), 
\end{align} 
\end{subequations}
then $S^1$ compactify the $\tau$-direction with the period $\beta_w$.
As a result, the momenta in the $\tau$-direction can be written as 
\begin{align}\label{sdqwfd}
2\pi n /\beta_w \equiv \omega_n,
\end{align} 
and it can be considered that we can  consider the gas 
at the temperature given as $\beta_w^{-1}$ 
in the space given by the space part of (\ref{qdjes}).
We write $\phi_\alpha$ ($\alpha=1,2$) as
\begin{align}\label{dbduy}
\phi_{\alpha}(\tau,r,\theta,\phi) 
= \frac{1}{\beta_w} \sum_{n=-\infty}^\infty
\varphi_{\alpha,n} (r,\theta,\phi) e^{-i\omega_n \tau}, \quad (\alpha=1,2)
\end{align} 
As a result, $Z$ in (\ref{vboqe}) can be written as
\begin{align}\label{vesvdtn}
Z= 
\int \! {\cal D} \phi_1 {\cal D} \phi_2 
\exp \Big[ \! 
-\frac{1}{2} \int \! d^3 x \! \int_0^{\beta} \! d\tau \,\gamma_w \, 
\left( \! \begin{array}{cc} \phi_1 & \phi_2 \end{array} \! \right) \! 
\left( \! 
\begin{array}{cc} 
\hat{G}_w+M_w^2 & -2ig^{\tau\tau} \mu_w \, \partial_\tau \\ 
2ij^{\tau\tau} \mu_w \, \partial_\tau & \hat{G}_w+M_w^2 
\end{array} 
\! \right) \!
\left( \! \begin{array}{c} \phi_1 \\ \phi_2 \end{array} \! \right) 
\Big],
\end{align}
where 
$ \hat{G}_w \equiv \, -j^{\tau\tau} \partial_\tau \partial_\tau + \gamma_w^{-1} j^{ij} \partial_i (\gamma_w \partial_j)$ and 
$M_w^2 \equiv m^2 - g^{\tau\tau}\mu_0^2$.
From the $M_w^2$ above, we can see that the upper bound of the value of the chemical potential 
in the case of the traversable wormhole at finite temperature is 
\begin{align}\label{dfrwyj}
\mu_w^c=m.
\end{align}

\subsection{Effective potential in the vicinity of the throat}
\label{qwpspw} 

In the same way with (\ref{jhfk}), putting $\phi_\alpha$ ($\alpha=1,2$) as
\begin{subequations}
\begin{align} 
\phi_\alpha &\equiv \sqrt{2} \, \alpha \, \cos \Theta + \tilde{\phi}_\alpha, \\
\phi_\alpha &\equiv \sqrt{2} \, \alpha \, \sin \Theta + \tilde{\phi}_\alpha, 
\end{align} 
\end{subequations}
then proceeding with the calculation from (\ref{vesvdtn}) in the same way with Sec.\ref{beytw}, we can obtain the following $Z$:
\begin{align}\label{aeofgj}
Z 
=
& 
\exp[-\alpha^2\beta_w V_3 M_w^2] 
\int\! {\cal D} \tilde{\phi}'_{1,n} {\cal D} \tilde{\phi}'_{2,n} \exp \bigg[ 
- \frac{1}{2\beta_w } \sum_{n=-\infty}^\infty \int_0^\infty \! dr \int_{-\infty}^\infty \! d^2 k_\perp \, \gamma_w \hat{G}_w (\tilde{\phi}_{1,n}^2 + \tilde{\phi}_{2,n}^2 )
\bigg],
\end{align}
where $V_3 \equiv \int dx^3 \gamma_w$. 
\newline

We switch our viewpoint from the entire region from $r=0$ to $\infty$ to only the vicinity of the throat. 
Therefore, replacing the $r$ in $\gamma_w \hat{G}_w$ in (\ref{aeofgj}) with $v+r$, 
we take $r$ up to $r^0$-order in $\gamma_w \hat{G}_w$ in (\ref{aeofgj}) as
\begin{align}\label{aspebj}
\gamma_w \hat{G}_w =
\frac{v^{5/2} \sin \theta}{\sqrt{2r}} (-\partial_\tau^2+M_w^2)-\frac{\sqrt{v} \sin \theta}{\sqrt{2r}}(v\partial_r -\hat{L}^2) +{\cal O}(r^0),
\end{align} 
where $\hat{L}^2$ is the squared angular-momentum operator:
\begin{align}\label{xcsili}
\hat{L}^2\equiv -(\frac{\partial_\theta (\sin\theta\partial_\theta)}{\sin\theta}+\frac{\partial_\phi^2}{\sin \theta^2}).
\end{align} 
Also, $V_3 $ is given as
\begin{align}\label{xhkd}
V_3=\frac{v^{5/2}\sin \theta}{\sqrt{2r}}+{\cal O}(r^0).
\end{align} 
In what follows, we proceed with our analysis by focusing on 
the vicinity of the throat, where the $r$ in what follows 
refers from $0$ to some infinitesimal number. 
\newline

We perform the expansion for the ($\theta,\phi$)-directions of $\phi_\alpha$ ($\alpha=1,2$) with the spherical harmonics as
\begin{align}\label{ovntyfd}
\tilde{\phi}_{\alpha,n}(r,\theta,\phi) 
= \frac{1}{\beta_{w}} 
\sum_{n=-\infty}^\infty 
\sum_{l,m} 
e^{-i\omega_n \tau}\tilde{\varphi}_{\alpha,n,l,m} (r) Y^l_m(\theta,\phi).
\end{align}
Then, proceeding with the calculation from (\ref{aeofgj}) by the same way with Sec.\ref{beytw}, we can obtain the free energy, $Z=\exp(-\beta_w F_w)$, as
\begin{align}\label{uvdpuil} 
F_w
=& \, 
\alpha^2 M_w^2 \frac{4\pi v^{5/2}}{\sqrt{2}}\int \frac{dr}{\sqrt{r}}
+
\frac{1}{\beta_w}
\sum_{n=-\infty}^\infty 
\sum_r 
\sum_{l,m}
\int_0^{M_w^2}d\Delta^2 H_{\Delta^2}(r,n,l),
\end{align} 
where $r$ in the summation takes all the real numbers from $0$ to $\infty$, and 
\begin{align}\label{caswfil} 
H_{\Delta^2}(r,n,l) \equiv \frac{1}{\omega_n^2-v\partial_r+l(l+1)+\Delta^2}.
\end{align} 

We finally obtain the critical temperature at some points where $r$ is near 0, 
so we take the contribution at each $r$ in (\ref{uvdpuil}) as 
\begin{align}\label{vepfil} 
\Gamma_w 
=& \, 
\alpha^2 M_w^2 \frac{4\pi v^{5/2}}{\sqrt{2r}}
+
\frac{1}{\beta_{w}}
\sum_{n=-\infty}^\infty 
\sum_{l,m}
\int_0^{M_w^2}d\Delta^2 H_{\Delta^2}(r,n,l), 
\end{align} 
where $r$ in (\ref{vepfil}) takes some values near $0$.

\subsection{$H_{\Delta^2}(r,n,l)$}
\label{acqdw} 

\begin{figure}[h!!!!!] 
\vspace{0mm} 
\hspace{0mm} 
\begin{center}
\includegraphics[clip,width=4.0cm,angle=-90]{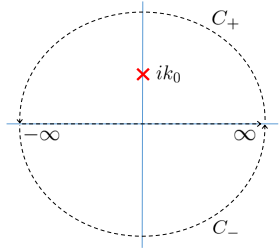} 
\end{center}
\vspace{0mm}
\caption{This figure represents the paths of $C^\pm$ and the position of the pole $ik_0$ 
to calculate $H_{\Delta^2}(r,n,l)$ in (\ref{vevrevo}) using the residue theorem.}
\label{svwef} 
\end{figure}

$H_{\Delta^2}(r,n,l)$ in (\ref{caswfil}) can be defined as the one to satisfy the following equation:
\begin{align}\label{vefio}
(\omega_n^2-v\partial_r+l(l+1)+v^2\Delta^2)H_{\Delta^2}(r,n,l)=\gamma_w^{-1}\delta(r) 
\end{align} 
where $\gamma_w^{-1}=\frac{\sqrt{2 r} }{v^{5/2}\sin \theta}+{\cal O}(r)$. 
Performing the Fourier expansion for the $r$-direction in $\tilde{D}_{\Delta^2}(r,n,l)$ and $\delta(r)$ as
\begin{align}\label{vevrevo}
H_{\Delta^2}(r,n,l)=\frac{1}{2\pi}\int_{-\infty}^\infty dk \tilde{J}_{\Delta^2} (k) e^{-ikr}, \quad \delta(r)=\frac{1}{2\pi}\int_{-\infty}^\infty dk e^{-ikr},
\end{align} 
it can be seen that $\tilde{J}(k)$ is given as 
\begin{align}\label{vklvcaw}
\tilde{J}_{\Delta^2}(k) &= \frac{\gamma_w^{-1}|_{r=r_0}}{iv(k-ik_0)} \quad(\textrm{$r_0$ is taken to 0 from the positive}), \nonumber \\
k_0 &\equiv \frac{1}{v}(l(l+1)+v^2\Delta^2+v^2\omega_n^2), 
\end{align} 
where $r_0$ is taken to some values near $0$ and 
$ik_0$ is located in the upper half-plane in the complex plane (see Fig.\ref{svwef}). 

Now we have obtained $\tilde{J}_{\Delta^2}(k)$ as in (\ref{vklvcaw}), we evaluate $\tilde{H}_{\Delta^2}(r,n,l)$ in (\ref{vevrevo}). 
We use the residue theorem for this, then we can see 
\begin{align}
&\bullet\quad 
\textrm{when $r > 0$, the contribution from the path of} 
\left\{
\begin{array}{l}
\textrm{the upper semi circle, $C_+$, is $\infty$},\\[1mm]
\textrm{the lower semi circle, $C_-$, is $0$},
\end{array}
\right.
\nonumber
\\
&\bullet\quad 
\textrm{when $r < 0$, the contribution from the path of}
\left\{
\begin{array}{l}
\textrm{the upper semi circle, $C_+$, is $0$},\\[1mm]
\textrm{the lower semi circle, $C_-$, is $\infty$},
\end{array}
\right.
\nonumber 
\\
&\bullet\quad 
\textrm{when $r = 0$, the contribution from the paths of $C_\pm$ are $\infty$.}
\nonumber 
\end{align} 
where $C_\pm$ and the position of $k_0$ are sketched in Fig.\ref{svwef}.
Finally, we can obtain as
\begin{align}\label{ksvsdw}
\tilde{H}_{\Delta^2}(r,n,l)
= e^{\frac{r}{v}(\Delta ^2+l(l+1)+\omega _n^2)} - \frac{\sqrt{2r_0}}{v^{7/2}\sin \theta}\Theta(-r) 
= e^{\frac{r}{v}(\Delta ^2+l(l+1)+\omega _n^2)}, 
\end{align} 
where $\Theta(-r)$ is the step function (it is 1 or 0 for negative or positive $r$), and our $r_0$ is some positive values near $0$. 
~\newline 

Now we have obtained $\tilde{D}_{\Delta^2}(r,n,l)$ as in (\ref{ksvsdw}). 
Then, in (\ref{vepfil}), performing the integration with regard to $\Delta^2$ , we can obtain the following $\Gamma_w$: 
\begin{align}\label{vsvhhuv} 
\Gamma_w 
=
\alpha^2 M_w^2 \frac{4\pi v^{5/2}}{\sqrt{2r}}+
\frac{v}{\beta_w r}
\sum_{n=-\infty}^\infty
\sum_{l,m}
(-e^{\frac{r}{v} (l(l+1)+\omega _n^2)} + e^{\frac{r}{v} (l(l+1)+M^2+\omega _n^2)}). 
\end{align} 

\subsection{Particle density and critical temperature in the vicinity of the throat} 
\label{desiuw} 

From (\ref{vsvhhuv}), according to $d_w=-d\Gamma_w/d\mu_w$, we can obtain the e.v. of the particle density as
\begin{align}\label{wobve}
d_w=
\frac{2\mu_w}{\beta_w}
\sum_{n=-\infty}^\infty
\sum_{l,m}
e^{\frac{r}{v} (l(l+1)+m^2-\mu_w^2+\omega _n^2)}, 
\end{align} 
where considering we close from some high temperatures to the critical temperature,
we have put $\alpha$ to 0 as well as Sec.\ref{vlsdkv} and \ref{vwbttdc}.
\newline

Now, we get the critical temperature in the vicinity of the throat. 
For this purpose, we first expand $d_w$ in (\ref{wobve}) around $r=0$ assuming $v$ is some finite number. 
Then, we apply $\mu_w^c=m$. At this time, the temperature is at the critical temperature, so we can write $\beta_w$ as $\beta_w^c$. 
Then, we can write (\ref{wobve}) as 
\begin{align}\label{nssd}
d_w=\frac{2m}{\beta^c_w}\sum_{n=-\infty}^\infty \sum_{l,m}
(1+\frac{2 m}{v}(l(l+1)+\omega _n^2)r)+{\cal O}(r^2/v^2). 
\end{align} 
Then, once treating the summations $\displaystyle \sum_{n=-\infty}^\infty \sum_{l,m}$ in (\ref{nssd}) 
as $\displaystyle \sum_{n=-N_0}^{N_0}\sum_{l=0}^{L_0}(2l+1)$, 
($N_0$ and $L_0$ are taken to infinity finally), 
we evaluate these summations. 
Then, 
we can obtain the critical temperature as 
\begin{align}\label{renre} 
T_w^c
&=\frac{d_w}{2 (L_0+1)^2 (2 m N_0+m)}\nonumber\\
&-\frac{ (2 \pi ^2 d^2 N_0(N_0+1)+3 L_0 (L_0+2) (L_0+1)^4 (2 m N_0+m)^2}{12 v(L_0+1)^6 (2 m N_0+m)^3}d_w \, r+{\cal O}(r^2/v^2),
\end{align} 
We can see that this $T_w^c$ goes to $0$ when $N_0$ and $L_0$ are sent to $\infty$, where 
\begin{itemize}
\item
Temperature $T_w$ is Euclid temperature as in around (\ref{qvsfm}),
$v$ is assumed as some finite number, and $r$ is measured from $v$ (see around (\ref{aspebj})),
\item
we have taken the part for small $r$ in the effective potential as in Sec.\ref{qwpspw}, 
which we consider to be effective for the phenomena in the vicinity of the throat. 
Then from the analysis of it, we have obtained the result as in (\ref{renre}).
\end{itemize}

Lastly, let us mention on the physical reason for the result (\ref{renre}), 
which may be considered that the dimension of the space becomes effectively 1 
in the vicinity of the throat (see Eq.(\ref{qdjes})). 
However, it depends on the coordinate system we take. 
Actually, the space can be 3 dimensional even in the vicinity of the throat, 
if we perform a coordinate transformation to $\zeta$ given as $r=v+\zeta^2/2v$. 
After all, the physical reason of the result (\ref{renre}) would be that contribution 
from the infrared region gets diverged for the curvature of the space in the vicinity 
of the throat, independently of the coordinate system we take.

Next, we comment on the coordinate dependence of the value of the critical temperature (\ref{renre}). 
Generally, the value of the Euclidean temperature would be changed up to the definition of the time 
coordinate. However, the result $T_w^c=0$ ($\beta_w^c=\infty$) at $r=0$ would not be changed 
as long as we do not consider some special transformation using some inverse of $r$ toward $t$. 
Therefore it would be considered the critical temperature is always 0 as long as some inertial systems are taken (see footnote in Sec.\ref{jeloiw}).

\section{
Phase structure of the normal/BEC states 
in the traversable wormhole space time 
and Josephson junction formed in vicinity of its throat}
\label{vewrc} 

\subsection{Phase structure}
\label{vewrc1} 

\begin{figure}[h!!!!!!!] 
\vspace{0mm} 
\hspace{0mm}
\begin{center}
\includegraphics[clip,width=6.0cm,,angle=-90]{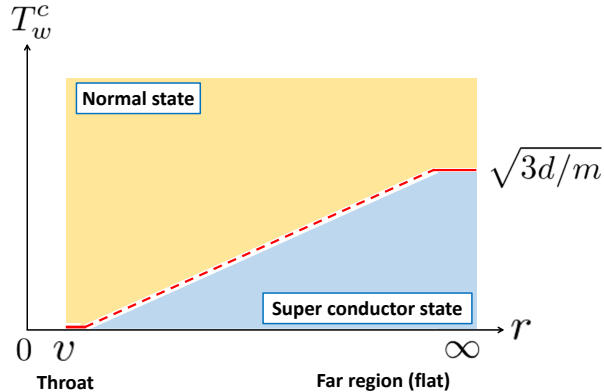} 
\end{center}
\vspace{0mm}
\caption{
Sketch of the phase structure for the critical temperature regarding BEC in the gas 
in the one side of the wormhole space time. {\bf  Left} and {\bf right red solid lines} respectively 
represent the critical temperatures in the vicinity of the throat and the far region that we have 
analytically obtained. On the other hand, {\bf red dotted line} is the interpolation between these. 
From this sketch, we can see that an analogous state to the Josephson junction can be formed 
at any temperatures in the vicinity of the throat.
}
\label{swefiohd}
\end{figure}

In the previous section, we have obtained the result that 
the critical temperature of BEC in the gas in the vicinity of the throat is 0 as in (\ref{renre}).
On the other hand, the far region of our wormhole space time (\ref{qdjes}) is asymptotically flat, 
and we have obtained the critical temperature of BEC in $D=3+1$ Euclid flat space time at finite temperature as in (\ref{dccile}). 
Summing up these results: 
\begin{itemize}
\item $T^c_w =0$ at $r \sim v$ (vicinity of throat),
\item $T^c_w=\textrm{(\ref{dccile})}$ at $r=\infty$ (far flat region).
\end{itemize}
Interpolating between these two results, we can sketch a phase structure like Fig.\ref{swefiohd}.

From Fig.\ref{swefiohd}, we can see that an analogous state to the Josephson junction is 
formed at any temperatures in the vicinity of the throat. 

Then, the question, whether the Josephson current is flowing or not in the vicinity of the throat, 
would arise, which we discuss in the next subsection. 

\subsection{On Josephson current to flow in the vicinity of the throat}
\label{vewrc2} 

The typical scales for the largeness of the Josephson junction and current in the laboratories would be roughly, 
\begin{align}\label{vvbsui}
\textrm{1[nm]-[$\mu$m] and 1[$\mu$A]-[mA].} 
\end{align}
Then, since Josephson current is a kind of the tunneling, 
it is considered that Josephson current would be more damped by the exponential 
as the width of the vicinity of the throat gets greater\footnote{
The ratio of the transmitted wave to the incident wave in the one-dimensional space with the potential barrier 
written in every textbook for the quantum mechanics is given as 
\begin{align}
\left(\frac{\textrm{Transmitted wave}}{\textrm{Incident wave}}\right)^2=\frac{1}{1+\frac{V_0^2 \sinh^2 \alpha L}{4\varepsilon(V_0-\varepsilon)}}\sim e^{-2\alpha L} 
\quad \textrm{for large $L$}, 
\end{align}
where $\varepsilon \le V_0$ and $\alpha= \sqrt{{2m}(V_0-\varepsilon)/{\hbar^2}}$, and 
$V_0$ and $L$ mean the height and width of potential barrier. 
$\varepsilon$ and $m$ mean the energy and the mass of the particles given as the wave function. 
}. 
Therefore, if the width of the normal state in the vicinity of the throat were larger than the scale in (\ref{vvbsui}), 
the Josephson current could not occur in practice.
Contrary, if it were in the scale of (\ref{vvbsui}), 
the Josephson current in the magnitude in (\ref{vvbsui}) might appear in the vicinity of the throat 
(though it is up to the property of the space time being in the normal state through which the Josephson current flows).

To answer this problem, we have to analyze the width of the normal state in the vicinity of the throat
and how much the wave function is damped when it tunnels. 
We cannot say any explicit things about this from the analysis in this paper. 

Normally considering, the wormhole is the astronomical object, 
therefore the width of the normal state is very larger than (\ref{vvbsui}).
Hence, the wave function of the Josephson current would be damped and vanish
when it goes through the normal state in the vicinity of the throat.

However, we could not exclude the possibility that it is not much damped for some effect of the curved space. 
Actually, there is a thought that the Hawking radiation is a kind of the tunneling \cite{Parikh:1999mf}. 
Hence, if the Hawking radiation exists, the Josephson current might also exist as the same tunneling phenomenology. 

One approach to this issue is to analyze the thermal de Broglie wavelength in the curved space time, 
which would be a future works.

\section{Summary}
\label{wvhou} 

In this study, we have investigated the phase structure for the BEC/normal states transition 
in the traversable wormhole space time filled by some gas to form BEC up to temperature. 

The first idea to make this work begin is something like the consideration we mentioned in Sec.\ref{jeloiw} , 
and we mention the points in the mechanism for the formation of BEC in this study in Appendix.\ref{dcmqt}.

As a result, we have obtained the result that the critical temperature of the gas for BEC is zero in the vicinity of the throat. 
Then, based on that result, we have pointed out that an analogous state to the Josephson junction is always 
formed in the vicinity of the throat. 

There is the problem of the exotic matter in the traversable wormhole space time as shown in Sec.\ref{vvbep}. 
This is a critical problem from the standpoint of the realizability of the Josephson junction in this study. 
For some references for this problem, see Sec.\ref{jeloiw}, and 
we in this study could care this problem by the consideration that 
it appears for some reason and can exist for some time, then this study 
is the one for that time. 
As for the analysis for how long time it can exist, it might come to some analysis for the unstable 
modes in the classical perturbations on the wormhole space time like the analysis for the Gregory-Laflamme instability.

If we could get these, we could reach the stage to discuss the realizability of the Josephson junction in this study, 
at which we should care the following 2 practical problems: 
1) Effect of the strong tidal force to the existence of our Josephson junction formed at the vicinity of the throat,
and 
2) how to actually create the situation where the space is filled by the gas.  
If we could finally clear these problems, we might consider our Josephson junction, realistically.  

The result in this study means that the state of the gas is changed from the normal state 
to the superconductor state at some point in the space. Investigating where it is and how 
it is are future works. In addition, the result in this study should be independent of the coordinate 
system we take as mentioned in the end of Sec.\ref{desiuw}. Checking it is also the future work. 
%
\newline

\begin{figure}[h!!!!!!!!!!] 
\vspace{0mm} 
\hspace{0mm}
\begin{center}
\includegraphics[clip,width=4.5cm,angle=0]{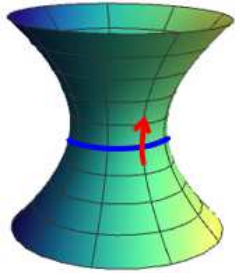} 
\end{center} 
\vspace{0mm} 
\caption{
This figure represents the image of the $ds_4^2$ part of the traversable wormhole space time 
given by (\ref{neqry}) to be used as the effective early cosmological model in the idea 
written in the body text (For how to obtain this figure, see \cite{nrvmwer}). 
The curved surface in this figure corresponds to the four-dimensional space time 
we exist; {\bf blue circle} represents its $S^3$ spatial space ($\Omega_3$-direction), and {\bf red direction} 
is $r$-direction and parameterizes the time development of that $S^3$ spatial space.
Here, {\bf throat part} is the $S^3$ space at $r=v$, which corresponds to the three-dimensional space 
at the beginning of the current cosmology. We can see it is not singular, therefore the space at the beginning 
of the cosmology is regularized. 
In this figure, $t_E$-direction is not included; it is originally the time-direction, 
however which is now being $S^1$ compactified into the imaginary direction and 
prescribes the temperature of the four-dimensional space depicted in this figure. 
Hence, the four-dimensional space depicted in this figure is supposed to be at some temperature. 
It would be interesting if the boundary between the superconductor and normal phases in our study 
can relate to Big Bang, where at this time, the superconductor region corresponds to the inflation era. 
} 
\label{rvvwr} 
\end{figure}

As one of the interesting examples in which our Josephson junction would play a 
very intriguing role, author considers some scalar-gravity system, where 
the scalar field is supposed to form BEC up to the temperature, c.f.~\cite{Kim:1997jf}, 
on the following Euclidean five-dimensional traversable wormhole space time 
($ds_4^2$ part is represented in Fig.\ref{rvvwr}):
\begin{eqnarray}\label{neqry}
ds_5^2 = dt_E^2 + ds_4^2, \quad \textrm{where} \quad ds_4^2 \equiv \frac{dr^2}{1-v^2/r^2}+r^2d\Omega_3^2,
\end{eqnarray}
as an effective model for the expanding early universe including the previous universe 
collapsing to the beginning of the current universe; the space time given by (\ref{neqry}) 
corresponds to the shape of the space time for the two universes joined by the throat part 
corresponding to the beginning of the current universe. 
Postponing the explanation for  (\ref{neqry}) to the next paragraph, 
we first say that it seems we could get the five-dimensional traversable wormhole space 
time as a solution regardless of the values of the cosmological constant if it comes 
to the five dimension (c.f.~\cite{Lemos:2003jb}). 
This is because the uniqueness theorem is the theorem in the four dimension.
 
In (\ref{neqry}), $S^3$-direction given by $r^2d\Omega_3^2$ corresponds to the three-dimensional 
spatial part we exist, and $r$-direction parameterizes the time development of that $S^3$ space.  
$t_E$-direction is the originally the time-direction, which is now being $S^1$ compactified 
into the imaginary direction and prescribes the temperature of the four-dimensional part of $ds_4^2$ 
(c.f.~\cite{DeBenedictis:2002wd}).

Hence, the space given by $ds_4^2$ part corresponds to the four-dimensional space time 
we exist, which is applied to the curved surface in Fig.\ref{rvvwr}, and is at some temperature  
determined by the period of the $t_E$-direction. The scalar field to form BEC up to the 
temperature also exists on the curved surface in Fig.\ref{rvvwr}.

The three-dimensional space at the beginning of the current universe corresponds 
to the throat part at $r=v$ in (\ref{neqry}), which is not singular, therefore it is considered 
that the space time at the beginning of the cosmology is regularized in this model. 
This point is one of the points in this model as the effective model for cosmology. 

Our Josephson junction is supposed to be formed in the vicinity of the beginning of the cosmology. 
It would be interesting if the boundary 
between the superconductor and normal phases in our study can relate to Big Bang, which 
is another point in this model (at this time, the superconductor region would correspond to the inflation era).

In conclusion, supposing we could get the five-dimensional traversable wormhole space time as 
a solution, it would be interesting to examine whether or not the effective model above can reproduce 
the picture of the  early cosmology described by the standard cosmology and give some  solutions 
for the unsettled problems in our current cosmology (c.f.~\cite{Roman:1992xj}).
\vspace{-0mm}

\appendix 

\section{Derivation of (\ref{sdwek})} 
\label{dgwfdn} 

The second kind modified Bessel function can be written using the first kind modified Bessel function $J_{\alpha}(x)$ as \cite{wikiBessel}
\begin{eqnarray}
K_\alpha(x) =\frac{\pi}{2} \frac{i^\alpha J_{-\alpha}(ix)-i^{-\alpha}J_{\alpha}(ix)}{\sin (\alpha \pi)}.
\end{eqnarray}
Using this, it can be written as
\begin{eqnarray}\label{sdaps}
&&
\int^\infty_0 \frac{1}{x}K_\alpha(x)K_\beta(x) dx 
=
\frac{i\left({\pi}/{2}\right)^2}{\sin (\alpha \pi)\sin (\beta \pi)}
\int^\infty_0 
\frac{dy}{y} 
({\cal A}_1+{\cal A}_2
)
,\\[1mm]
&&{\cal A}_1= i^{\alpha+\beta}J_{-\alpha}(y)J_{-\beta}(y)+i^{-\alpha-\beta}J_{\alpha}(y)J_{\beta}(y),\nonumber\\[1mm]
&&{\cal A}_2=-i^{\alpha-\beta}J_{-\alpha}(y)J_{\beta}(y)-i^{-\alpha+\beta}J_{\alpha}(y)J_{-\beta}(y).\nonumber
\end{eqnarray}
We can check ${\cal A}_1+{\cal A}_2=0$ for $\alpha\not=\beta$, therefore (\ref{sdaps}) is 0 for $\alpha\not=\beta$. 
\newline

Next, for $\alpha=\beta$, 
performing the Wick rotation as $\alpha \to i\alpha$, the part of ${\cal A}_1$ can be calculated as
\begin{eqnarray}\label{slowd}
-
\frac{i\left({\pi}/{2}\right)^2}{\sin^2 (i\alpha \pi)}
\int^\infty_0 
\frac{dy}{y} {\cal A}_1
= 
-\frac{\left({\pi}/{2}\right)^2}{\alpha \sinh (\alpha \pi)}.
\end{eqnarray}
Putting $\beta$ as $\beta \to \alpha+\Delta \alpha$ where $\Delta \alpha$ is taken to $0$ finally, 
the part of ${\cal A}_2$ can be calculated as
\begin{eqnarray}\label{ksdc}
&&
-\frac{2i(\pi/2)^2}{\sin (\alpha \pi)^2}
\int^\infty_0 
\frac{dy}{y} {\cal A}_2
=\frac{i \pi^2}{2 \alpha + \Delta \alpha} 
\frac{1}{\sin (\alpha \pi)} 
\frac{1}{\Delta \alpha},
\end{eqnarray}
where we have used the formula 
$
\int_0^\infty 
\frac{dz}{z}
J_\alpha (z) J_\beta (z)
=
\frac{2}{\pi}
\frac{ \sin (\frac{\pi}{2} (\alpha-\beta)) }{ \alpha^2- \beta^2}
$ \cite{wikiBessel}. 
Putting $\alpha \to i \alpha$ and $\Delta\alpha \to i \Delta\alpha$, 
\begin{eqnarray}\label{npwe}
(\ref{ksdc})=
\frac{\pi^2}{2\alpha \sinh (\alpha \pi)} 
\frac{1}{\pi \Delta \alpha}.
\end{eqnarray}
~\newline

Summarizing the results above,
\begin{eqnarray}\label{acwe}
(\ref{sdaps})= \left\{
\begin{array}{ll}
0 & \textrm{for $\alpha \not= \beta$}, \\
[1mm]
{\displaystyle 
\frac{\pi^2}{2\alpha \sinh (\alpha \pi)} 
\frac{1}{\pi \Delta \alpha}
-\frac{\left({\pi}/{2}\right)^2}{\alpha \sinh (\alpha \pi)}
}
& \textrm{for $\alpha = \beta$}.
\end{array}
\right.
\end{eqnarray} 
When $\Delta \alpha \to 0$, the results of each case in (\ref{acwe}) can be written at once as
\begin{eqnarray}
\int^\infty_0 \frac{1}{x}K_\alpha(x)K_\beta(x) dx = \frac{\pi^2\delta(\alpha-\beta)}{2\alpha \sinh (\alpha \pi)}, 
\end{eqnarray}
where we have regarded $\pi \Delta \alpha$ in (\ref{acwe}) as $dx$, 
then regarded it as $\delta(\alpha-\beta)$ 
(generally, the delta function $\delta(x)|_{x=0}$ is equivalent to $dx^{-1}$, where $dx$ is the one in 
$
\int dx \delta(x)=dx \delta(x)|_{x=0}
$).

\section{Mechanism for the formation of BEC in this study}
\label{dcmqt} 


In this appendix,  the technical points in the mechanism of the formation of BEC in this study is mentioned.

First of all, the fundamental thought in our model could be considered 
as the one in the usual fundamental model of BEC like the one given 
in \cite{Kapusta:2006pm}, and what we have done is to apply it to curved 
space times. Therefore, the fundamental form of our Hamiltonian is 
given by the one in the grand canonical ensemble, $H-\mu N$. 

Then, if naively considering, it appears that we may be able to move only the chemical potential, only the number of particles 
or only temperature without the rests of these not moved. However, this is wrong. 
The structure in the microscopic states in the ground canonical ensemble is very complex, 
and these are closely related each other as can be seen from the Bose distribution function 
(meaning of variables are mentioned in the following),
\begin{eqnarray}\label{cwcrp}
\langle n_r \rangle=\frac{1}{e^{\beta(e_r-\mu)}-1}.
\end{eqnarray}

Therefore, this appendix begins from the derivation of (\ref{cwcrp}) 
as confirmation. Then based on it, the technical points in the mechanism 
of the formation of BEC in this study is mentioned.

\subsection{Bose distribution function: Relation that number of particles increased when chemical potential increased}
\label{nwvlt} 

First, in order to obtain the form of the Hamiltonian in the ground canonical ensemble, 
let us suppose that there are $M$ copied systems, then suppose $M_{N,i}$ as the number 
of the systems having the energy value specified by $N$ and $i$, $E_{N,i}$, and containing 
$N$ particles. These $M_{N,i}$, $E_{N,i}$ and $N$ get the following constraints:
\begin{eqnarray}\label{vnelrv}
\sum_{N,i} M_{N,i}=M, \quad 
\sum_{N,i} E_{N,i} M_{N,i}=E_0, \quad 
\sum_{N,i} N M_{N,i}=N_0,
\end{eqnarray}
where 
$\sum_{N,i}$ means 
$\sum_{N=0}^{N_0} \sum_{i}$,
and 
$M$, $E_0$ and $N_0$ are constants.

Now, consider the $\Gamma$ space (the space that has canonical 
variables of each of $N_0$ particles in the whole of $M$ copied 
systems as its coordinates). 
Then, each infinitesimal region in the $\Gamma$ space corresponds 
to a set of $M_{N,i}$, and if we check each infinitesimal region of some 
region, we would find that there are a number of the same sets $M_{N,i}$ 
(for the meaning of ``the same'', read it out from the $W$ in (\ref{dvcwb})). 
Here, let us suppose the Ergodic hypothesis (probability that each $M$ 
copied system takes some one of $M_{N,i}$ states is always the same 
in the region determined by $N_0$ and $E_0$) is held in the $\Gamma$ space. 
As a result, the most appearing ``the same'' set of $M_{N,i}$ is considered 
to be the set for the thermal equilibrium state. 
Therefore, let us obtain the most appearing set of $M_{N,i}$ in the region 
determined by $N_0$ and $E_0$ in the $\Gamma$ space. This problem 
is to obtain the set of $M_{N,i}$ which maximizes the following $W$: 
\begin{eqnarray}\label{dvcwb}
W=\frac{M!}{\Pi_{N,i}M_{N,i}!}. 
\end{eqnarray}
Then, using the method of Lagrange multiplier and Stirling's approximation, 
we can finally express all such the $M_{N,i}$ at once as
\begin{eqnarray}
\label{nccwc}
\frac{M_{N,i}}{M} \!\! &=& \!\! \frac{1}{\Xi}\exp[-\beta(E_{N,i}-N \mu)],\\ [1.0mm]
\label{nxvwu}
\Xi \!\! &=& \!\! \sum_{N,i}\exp[-\beta(E_{N,i}-N \mu)],
\end{eqnarray} 
where $M$, $E_0$ and $N_0$ are supposed to be very large positive integers to use 
Stirling's approximation, and $\beta$ and $\mu$ mean the inverse temperature and 
chemical potential. (\ref{nccwc}) means the probability that the system with $N$ particles 
and the energy $E_{N,i}$ appears. 

From here, let us suppose that the particles in the system follow the Bose statistics. 
As a result, the energy values are discretized and the number of particles to take each 
energy value is no limited. Then we can rewrite and replace as
\begin{eqnarray}
N=\sum_{r=\textrm{lowest}}^{\textrm{highest}} n_r \quad {\rm and} \quad
\sum_{N,i} \to \sum_{N=0}^{N_0} \sum_{\{n_r\}}, \quad 
E_{N,r} \to \sum_{r=\textrm{lowest}}^{\textrm{highest}} e_r n_r.
\end{eqnarray}
where ``lowest'' and ``highest'' mean those of the discretized energy levels, ``$\sum_{\{n_r\}}$'' 
means to produce all the sets of $n_r$ satisfying $N=\sum_r n_r$, and $e_r$ mean the values 
of the energy labeled by $r$ that each particles takes. Therefore, finally $\Xi$ in (\ref{nxvwu}) 
can be given as
\begin{eqnarray}
\label{leqfks}
\Xi 
=\sum_{N=0}^{N_0} \sum_{\{n_r\}} \exp[-\beta\sum_r\varepsilon_r n_r]
=\sum_{N=0}^{N_0} \sum_{\{n_r\}}\prod_r \exp[-\beta\varepsilon_r n_r],
\end{eqnarray}
where $\varepsilon_r \equiv e_r-\mu$ and $\sum_r$ means $\sum_{r=\textrm{lowest}}^{\textrm{highest}}$. 
Then, it is known that $\sum_{N=0}^{N_0} \sum_{\{n_r\}}$ can be treated as $\prod_r \sum_{n_r}$, namely 
we can independently perform the summation for each $n_r$ in the range $\sum_r n_r \le N_0$. At this 
time, if $N_0$ is infinity, $\sum_{N=0}^{N_0} \sum_{\{n_r\}}$ can be treated as $\prod_r \sum_{n_r=0}^\infty$. 
Therefore, supposing that $N_0$ is infinity, 
\begin{eqnarray}\label{lbdghs}
\Xi \!\!\! &=& \!\!\! 
\prod_r \sum_{n_r=0}^\infty \exp[-\beta\varepsilon_r n_r].
\end{eqnarray}
At this time, if $\exp[-\beta\varepsilon_r]<1$, namely if $\varepsilon_r>0$, for all $r$, 
\begin{eqnarray}\label{ccwoe}
\Xi=\prod_r (1-\exp[-\beta\varepsilon_r ])^{-1}. 
\end{eqnarray}
On the other hand, if any one of $\varepsilon_r \le 0$, $\Xi$ gets diverged.

Now, we obtain the e.v. of $n_r$, which can be written as 
\begin{eqnarray}
\langle n_r \rangle=
\frac
{\sum n_r \exp [-\beta\sum_r\varepsilon_r n_r]}
{\sum \exp [-\beta \sum_r\varepsilon_r n_r]}, 
\end{eqnarray}
where $\sum$ above mean $\sum_{N=0}^{N_0} \sum_{\{n_r\}}\prod_r$. 
Since $\langle n_r \rangle=-\frac{1}{\beta}\frac{\partial \ln \Xi}{ \partial \varepsilon_r}$, 
therefore using (\ref{ccwoe}), we can obtain $\langle n_r \rangle$ given in (\ref{cwcrp}),
and we can get a key fact in the formation of BEC in this study, 
the number of the particles is increased when the chemical potential is increased.

\subsection{
Keep density of gas constant in decreasing temperature, 
and for this purpose, increase chemical potential, 
then finally formed BEC described by the constants}
\label{nlfpret} 

As the fundamental thought in our model, we want to keep the number of the particles 
of the gas as it is when we decrease the temperature. Then, since there is a relation (\ref{cwcrp}), 
the chemical potential should be raised.

However, as can be seen from the text under (\ref{ccwoe}), there is the upper limit 
for the value the chemical potential can take. Actually, (\ref{owvu}), (\ref{dsvuje}) 
and (\ref{dfrwyj}) would be that.

Therefore, the system needs to take some way other than the growing up of the chemical potential 
when the chemical potential reaches the upper limit, which is to have the $\alpha$ in (\ref{jhfk}) have 
some finite value. This is the definition of the formation of BEC.


\begin{thebibliography}{100}
\bibitem{nrvmwer} 
Fig.\ref{rvvwr} has been obtained based on the program in ``Trying to plot a wormhole; getting bad results'' in mathematica.stackexchange.com
\end{thebibliography}
\end{document}